\documentclass[aps,prl,twocolumn,superscriptaddress,groupedaddress]{revtex4}  

\usepackage{orcidlink}
\usepackage[T1]{fontenc}

\input{definitions}

\usepackage{cancel}
\usepackage{soul}
\usepackage[normalem]{ulem}

\begin{document}
\begin{flushright}
\end{flushright}
\title{Search for an Invisible Z$^\prime$ in a Final State with Two Muons and Missing Energy at Belle II}

  \author{I.~Adachi\,\orcidlink{0000-0003-2287-0173}} 
  \author{K.~Adamczyk\,\orcidlink{0000-0001-6208-0876}} 
  \author{L.~Aggarwal\,\orcidlink{0000-0002-0909-7537}} 
  \author{H.~Ahmed\,\orcidlink{0000-0003-3976-7498}} 
  \author{H.~Aihara\,\orcidlink{0000-0002-1907-5964}} 
  \author{N.~Akopov\,\orcidlink{0000-0002-4425-2096}} 
  \author{A.~Aloisio\,\orcidlink{0000-0002-3883-6693}} 
  \author{N.~Anh~Ky\,\orcidlink{0000-0003-0471-197X}} 
  \author{D.~M.~Asner\,\orcidlink{0000-0002-1586-5790}} 
  \author{T.~Aushev\,\orcidlink{0000-0002-6347-7055}} 
% ...
\collaboration{The Belle II Collaboration}

\begin{abstract}
    \vspace*{20pt}
    The \lmultau\  extension of the standard model predicts the existence of a lepton-flavor-universality-violating \zprime\ boson that couples only to the heavier lepton families. We search for such a \zprime\ through its invisible decay in the process $e^+ e^- \to \mu^+ \mu^- Z^{\prime}$.
We use a sample of electron-positron collisions at a center-of-mass energy of 10.58~GeV  collected by the Belle~II experiment in 2019--2020, corresponding to an integrated luminosity of 79.7~fb$^{-1}$.
We find no excess over the expected standard-model background. We set 90$\%$-confidence-level upper limits on the cross section for this process as well as
on the coupling of the model, which ranges 
from $3 \times 10^{-3}$ at low $Z^\prime$ masses  to 1 at $Z^\prime$ masses of 8~\gevcc.

\end{abstract} 
\maketitle

Recent experimental observations are in tension with the standard model~(SM) of particle physics. 
A notable example
is the difference between the measured and expected values of the muon anomalous magnetic moment~\cite{PhysRevD.73.072003, PhysRevLett.126.141801}.
In addition, the SM is known to provide an incomplete description of nature since, among other prominent issues, it does not address the phenomenology related to the existence of dark-matter~\cite{BERTONE2005279}, specifically  the prediction of the  observed relic density.
A simple
way to explain both phenomena is the \lmultau\ extension of the SM~\cite{PhysRevD.43.R22,Shuve:2014doa, Altmannshofer:2016jzy}.
This model gauges the difference of the muon and tau lepton numbers, giving rise to a massive, electrically neutral, vector boson, the \zprime. 
This particle would couple to the SM only through  $\mu$, $\tau$, $\nu_\mu$, and $\nu_\tau$ with  coupling  $g^{\prime}$. 
The  
\zprime, with such a lepton-flavor-universality-violating coupling, would contribute to the muon magnetic moment 
and, for certain values of $g^{\prime}$ and mass $M_{\rm{Z}^\prime}$, would explain the observed anomaly~\cite{PhysRevLett.113.091801}. 
This model may resolve the tensions in flavor observables reported by the LHCb, Belle and  \textit{BABAR} collaborations~\cite{LHCb:2021trn, BaBar:2013mob, Belle:2015qfa, LHCb:2015gmp, Belle:2017ilt, LHCb:2017rln, Belle:2019rba, Crivellin:2022obd, SALA2017205, Chen:2017usq, Greljo:2021xmg}.
It may also reproduce the observed dark-matter relic density, assuming dark-matter is charged under \lmultau. Two possible scenarios have been proposed, suggesting sterile neutrinos~\cite{Shuve:2014doa} or light Dirac fermions~\cite{Altmannshofer:2016jzy} as  dark-matter candidates.

In this Letter we report a search, performed with the Belle~II experiment, for the \zprime\ in the process $e^+e^- \to \mu^+\mu^- \rm{Z}^{\prime}$ with Z$^{\prime} \to \text{invisible}$, where the \zprime\ is radiated off a muon. We consider two different scenarios.
If the \zprime\ couples only to SM particles, a model henceforth referred to as the ``vanilla'' \lmultau\  model, the invisible decay happens only through neutrinos, with a branching fraction $\cal{B}\it(\rm{Z}^{\prime} \to \nu \bar{\nu})$ that varies  between $\sim$33\% and $\sim$100\% depending on the\zprime\ mass~\cite{PhysRevD.95.055006, hunting}.
Alternatively, the \zprime\ can decay directly into a pair of dark-matter particles $\chi \bar{\chi}$ with a coupling constant $\alpha_{D} = {g^{\prime}_{D}}^2/ {4\pi}$, and there is no \textit{a priori} reason for $\alpha_{D}$ to be small. In this case, one can expect
$g^\prime_{D} \gg g^\prime$, which implies  $\cal{B}\it(\rm{Z}^{\prime} \to \chi \bar{\chi}) \approx \rm1$: we henceforth refer to this second scenario as the ``fully invisible'' \lmultau\ model.

We provide  results for each of the two scenarios for \zprimemass~$<$~9~\gevcc. In the vanilla model, the intrinsic width $\Gamma_{\rm{Z}^\prime}$ of the Z$^\prime$ is negligible compared with the experimental resolution. In the fully invisible model, $\Gamma_{\rm{Z}^\prime}$ depends on \alphad: it is negligible  for  values of \alphad\ smaller than 1 for \zprimemass\ $\approx$ 1.5 \gevcc, and smaller than 0.1 for \zprimemass\ $\approx$ 4.5 \gevcc. We focus our analysis on the case in which $\cal{B}\it(Z^{\prime} \to \chi \bar{\chi}) \approx \rm1$ and $\Gamma_{Z^\prime}$ is negligible. We study separately one example in which $\Gamma_{Z^\prime}$ is not negligible and assume one benchmark value such that $\Gamma_{Z'}=0.1 M_{Z'}$, corresponding to  $\alpha_{D} = 2.9$.

Searches for a Z$^{\prime}$ decaying to muons have been performed by the \textit{BABAR}~\cite{TheBABAR:2016rlg}, Belle~\cite{Belle:2021feg}, CMS~\cite{cms} and ATLAS~\cite{ATLAS:2023vxg} experiments: they only  constrain the vanilla \lmultau\ model in the parameter space we explore. Searches for an invisibly decaying Z$^{\prime}$ have been performed 
by the NA64-$e$ experiment~\cite{Andreev:2022txy} in the low-mass region and by Belle~II using data collected during the commissioning run in 2018, with a luminosity of 0.276\invfem~\cite{PhysRevLett.124.141801}: these searches set constraints both in the vanilla and fully invisible \lmultau\ models. 

We  use a  sample of $e^+e^-$~collisions collected by Belle~II at the center-of-mass (c.m.) energy of the $\Upsilon$(4S) resonance, $10.58\gev$, in 2019--2020, corresponding to a total integrated luminosity of 79.7\invfem~\cite{lumi}. 
This search supersedes that in Ref.~\cite{PhysRevLett.124.141801}, with an integrated luminosity nearly 300 times larger, improved muon identification, and the use of refined analysis algorithms.

The invisible-\zprime\ signature is a narrow enhancement in the  distribution of the mass \mrec\ of the system recoiling against a muon pair. In the following, recoil quantities are computed by using the measured muon momenta and the knowledge of the initial-state total momentum.
These quantities coincide with \zprime\ properties for signal events and typically correspond to undetected SM
particles for background. 
We use the recoil mass squared \mrecSquare\ since this quantity has a  smoother distribution than \mrec\ for low masses. 
We select events with exactly two charged particles identified as muons, and negligible additional activity in the detector. 
The dominant backgrounds are processes which produce two muons and missing energy.
These are primarily $e^+e^- \rightarrow \mu^+\mu^-(\gamma)$ events with one or more undetected photons, $e^+e^- \rightarrow \tau^+\tau^-(\gamma)$ events with both $\tau$ leptons decaying to muons and neutrinos, and $e^+e^- \rightarrow e^+e^-\mu^+\mu^-$ events (dominated by two-photon fusion production) with electrons outside the detector acceptance.

We extract the signal yield from a fit to the two-dimensional distribution of $M^2_{\rm{recoil}}$ and  the polar angle $\theta_{\rm{recoil}}$ of the recoil momentum with respect to the detector axis.
Control samples are used to check simulation predictions and to infer correction factors. Selections  are optimized using simulated events prior to examining data. 
However, one of the corrections based on control samples was derived after observing a discrepancy in the data with respect to the simulation.

The Belle~II detector~\cite{Abe:2010sj} operates at the interaction region of the SuperKEKB electron-positron collider~\cite{superkekb}, located at KEK in Tsukuba, Japan. The energies of the electron and positron beams are $7\gev$ and $4\gev$
with a boost  of the c.m.\ frame $\beta\gamma = 0.28$ relative to the laboratory frame. 
The detector consists of several subdetectors arranged around the beam pipe in a cylindrical structure.
Subdetectors relevant for this analysis are briefly described here in order from innermost out; more details are given in Refs.~\cite{Abe:2010sj, ref:b2tip}.
The innermost component is the vertex detector, consisting of two inner layers of silicon pixels and four outer layers of silicon strips. The second pixel layer is partially installed, covering one sixth of the azimuthal angle.
The main tracking subdetector is a large helium-based small-cell drift chamber.
An electromagnetic calorimeter (ECL) consists of a barrel and two endcaps made of CsI(Tl) crystals.
A superconducting solenoid provides a 1.5~T magnetic field.
A $K^0_L$ and muon subdetector  is made of iron plates providing the magnetic flux-return yoke, alternated with resistive-plate chambers and plastic scintillators in the barrel, and with plastic scintillators only in the endcaps.
The longitudinal and transverse directions, and polar angle $\theta$ are defined with respect to the detector's cylindrical axis in the direction of the electron beam. 
In the following, quantities are defined in the laboratory frame unless otherwise specified.

Particle identification is implemented through the definition of likelihoods for each charged particle hypothesis by combining information from all the subdetectors.
Identification of muons  relies mostly on charged-particle penetration depth in the muon detector for momenta larger than 0.7~GeV/$c$ and on information from the drift chamber and ECL otherwise.
The ratio  between the muon likelihood and the sum of the likelihoods of all particle hypotheses is required to be greater than 0.5. This retains 93--99\% of muons, and  rejects 80--97\% of  pions, depending on their momenta. 
Electrons, used in control-sample studies,  are identified primarily by comparing momenta
with energies of associated ECL depositions, with a similar likelihood-ratio method.
Photons are reconstructed from ECL depositions with energy greater than 100~MeV that are not associated with any track.

Signal events are simulated 
using \texttt{MadGraph5\textunderscore aMC@NLO 2.6.6}~\cite{Alwall2014} with initial state radiation. 
The signal \mrecSquare\ resolution ranges from a minimum of 0.06~\gevsq at 80~\gevsq to a maximum of 0.23~\gevsq at 9~\gevsq.
We generate 582\zprime\ samples, with negligible $\Gamma_{Z^\prime}$,  corresponding to mass hypotheses ranging from 0.01 to $8.5\gevcc$ in steps of  
3 to 202 \mevcc, following the resolution. 
Background events are simulated using the following generators: 
$e^+e^- \rightarrow \mu^+\mu^-(\gamma)$ with \texttt{KKMC 4.19}~\cite{ref:kkmc},
$e^+e^- \rightarrow \tau^+\tau^-(\gamma)$ with \texttt{KKMC 4.19}~\cite{ref:kkmc} in combination with \texttt{TAUOLA 3.1}~\cite{ref:tauola},
$e^+e^- \rightarrow e^+e^-\mu^+\mu^-$ and  
$e^+e^- \rightarrow e^+e^-e^+e^-$ with \texttt{AAFH}~\cite{ref:fourlepton},
$e^+e^- \rightarrow \pi^+\pi^-(\gamma)$ with \texttt{PHOKHARA 9.1}~\cite{ref:phokhara},
 and $e^+e^- \rightarrow e^+e^-(\gamma)$ with \texttt{BabaYaga@NLO}~\cite{ref:babayaga}. 
Backgrounds coming from the final states $q\bar q$ ($q=u,d,s,c,b$), $J/\psi\gamma$, $\psi(2S)\gamma$ with $J/\psi, \psi(2S) \rightarrow \mu^+\mu^-$, and $\mu^+\mu^- \nu \bar\nu$
are found to be negligible.
Detector geometry and interactions of final-state particles with the detector material are simulated using \texttt{\textsc{Geant4 10.06}}~\cite{ref:geant4} and the Belle~II Analysis Software Framework~\cite{basf2, basf2-zenodo}.

The search uses an online event selection (trigger) that requires events with at least one pair of tracks in a restricted polar-angle acceptance,   $\theta\in[37,120]^\circ$, and an azimuthal opening angle larger than $90^\circ$ or $30^\circ$ for data collected in 2019 and 2020, respectively (two-track trigger). 
A dedicated trigger veto rejects events consistent with Bhabha scattering.

In the offline analysis,  we require that tracks originate from the interaction point, with transverse and longitudinal projections of their distance of closest approach smaller than 0.5 and 2.0~cm, respectively, to reject spurious and beam-induced background tracks.
We require events to have exactly two oppositely charged particles identified as muons, that pass the trigger requirements and have transverse momenta larger than 0.4 GeV/$c$.
We reject events with opening angles between the muons in the c.m.\ frame larger than $179^\circ$,
to suppress $\mu^+\mu^-(\gamma)$ backgrounds, which typically produce back-to-back muons.
For \mrecSquare\ $<4$~\gevsq, most of the $\mu^+\mu^-(\gamma)$ background comes from  events with  single-photon emission:  we require \trec\ to be within the ECL barrel acceptance $[34,123]^\circ$ so as to exclude regions where photons can escape undetected. 
Additionally, we reject events with  $\theta_{\rm{recoil}}\in[89,91]^\circ$, where  there is a 1.5 mm-wide gap in the ECL instrumentation.
For \mrecSquare\ $>4$~\gevsq, $\mu^+\mu^-(\gamma$) background arises predominantly from
multiple photon emission. In this case the recoil direction does not coincide with the direction of the lost photons, so we instead require $\theta_{\rm{recoil}} < 123^\circ$ because the signal is dominantly produced in the forward direction due to the c.m. boost. 
To suppress $\mu^+\mu^-(\gamma)$ backgrounds, we impose a photon veto: we require the total energy of all photons to be less than 0.5~GeV and no photon to be within $15^\circ$ of the recoil momentum.
To suppress  $\mu^+\mu^-(\gamma)$ and $e^+e^-\mu^+\mu^-$ backgrounds, we require that the transverse recoil momentum in the c.m.\ frame exceed 0.5~GeV/$c$.

After these selections, the remaining background comes dominantly from $\tau^+\tau^-(\gamma)$ events with $\tau$ leptons decaying to muons or to pions misidentified as muons in the region \mrecSquare~$<$ 50 \gevsq, 
and from $e^+e^-\mu^+\mu^-$ events elsewhere. The background from $e^+e^- \to \mu^+\mu^-(\gamma)$ is subleading across the entire mass range.

The final selection uses an artificial neural network, denoted as Punzi-net~\cite{punzi-loss}, trained on simulated signal and background events, and specifically designed to optimize a figure of merit~\cite{Punzi:2003bu}  for all \zprime\ mass hypotheses simultaneously. 
We use  as inputs the  four kinematic variables,  all defined in the c.m.\ frame, with the highest discriminating power: the transverse momentum of 
a muon with respect to the dimuon thrust axis~\cite{BRANDT196457, PhysRevLett.39.1587}; the transverse momentum 
of the higher-energy muon with respect to the momentum direction of the lower-energy muon; the longitudinal momentum 
of the higher-energy muon with respect to the momentum direction of the lower-energy muon; and the transverse momentum of the dimuon system. 
The first three variables exploit mostly the kinematic properties of \zprime\ production through radiation from a final state muon, compared with $\tau^+\tau^-(\gamma)$ events, in which the recoil momentum arises from neutrinos from $\tau$ decays. The fourth variable exploits the kinematic features of $\mu^+\mu^-(\gamma)$ and $e^+e^-\mu^+\mu^-$ backgrounds, which typically have  low transverse momenta. The Punzi-net produces an output between 0 (background) and 1 (signal): we select events with an output larger than 0.5.
Additional details 
are given in Ref.~\cite{punzi-loss}.

The resulting signal efficiency is typically 5\%, nearly uniform as a function of \mrecSquare. 
The Punzi-net selection  removes nearly all $\tau^+\tau^-(\gamma)$ background for \mrecSquare~$<$~50~\gevsq, with a sensitivity gain between 5 and 15, depending on the mass.
The residual background comes dominantly from  $\mu^+\mu^-(\gamma)$ in the region \mrecSquare~$<$~50~\gevsq and from a large irreducible contribution of $e^+e^-\mu^+\mu^-$  for \mrecSquare~$>$~50~\gevsq, where the Punzi-net selection has limited discriminating power.  

The region \mrecSquare~$<$~1~\gevsq\ is dominated by the $\mu^+\mu^-(\gamma)$ process with a single photon emission. Above 1~\gevsq\ the $\mu^+\mu^-(\gamma)$ process contributes mostly with events containing two radiated photons. Typically, one photon is collinear with the beams and outside the acceptance, while the other is emitted in the direction of one of the gaps between the barrel and the forward or backward ECL endcaps.  For \mrecSquare\ in the 1--50~\gevsq\ range, this produces two distinctive bands in the \treccm-$M^2_{\rm{recoil}}$ plane,
 where \treccm\ is the polar angle of the recoil momentum in the c.m.~frame. 
This feature is exploited in a two-dimensional fitting procedure, 
which incorporates the expected background shapes due to $\mu^+\mu^-(\gamma)$ events, doubling the sensitivity relative to a one-dimensional fit.

We fit the data by maximizing  a binned likelihood  based on signal and background two-dimensional templates obtained from simulation.
The parameter of interest is the signal cross section, with the background normalization determined by the fit.
The \mrecSquare\ bin widths vary across the spectrum and are set  to the signal \mrecSquare\ resolution. 
The binning in \treccm\  is determined by the distribution of $\mu^+\mu^-(\gamma)$ events and depends on \mrecSquare. The number of bins varies from one (for high \mrecSquare) to five  (for low \mrecSquare).
We perform a squared-mass scan in steps corresponding to one unit of signal \mrecSquare\ resolution, testing all simulated mass hypotheses. 
Each fit is performed in search windows of 20 \mrecSquare\ bins centered around 
each hypothesis.
Uncorrelated systematic uncertainties on the signal and background shapes (see below) are included in the respective templates by introducing in each bin of the template Gaussian nuisance parameters constrained by the corresponding uncertainties.   
A frequentist procedure based on the profile-likelihood ratio $\tilde{t}_{\mu}$~\cite{cowan2011} is used to obtain 90\% confidence-level (C.L.) intervals on the cross section.
We use the \texttt{pyhf} software package~\cite{Heinrich2021} for inference and check the consistency of our results with simplified simulated samples. 

We also consider the scenario in which  $\Gamma_{\rm{Z}'}$ is not negligible, as expected for large \alphad\ values~\cite{Crivellin:2022obd, Crivellin:2022gfu}, and 
study one benchmark case that
assumes $\Gamma_{Z'}=0.1 M_{Z'}$.
We account for the nonzero width in the fitting procedure by changing the shape of the signal templates to Breit-Wigner distributions with the widths $\Gamma_{Z'}$ convolved with  Gaussian resolution functions.
We use only one-dimensional \mrecSquare\ templates  and enlarge the search windows  to cover the sizable signal  width.

Three control samples are used to validate the analysis  and  estimate systematic uncertainties.
The $\mu\mu\gamma$ control sample is obtained by reversing the photon-veto criteria thus requiring 
a photon of energy greater than $1\gev$ within 15$^\circ$ of the recoil momentum direction. The $e\mu$ and $ee$ control samples are obtained by requiring one or both tracks to be identified as  electrons and then applying all the other  selection criteria.

The efficiency of the two-track trigger is studied with  the $e\mu$ control sample with events collected by an ECL-based trigger, which is used as a reference. 
This requires that the total energy deposition in the barrel and forward endcap exceed 1 GeV. The two-track trigger efficiency increases  from $89$ to $93 \%$ as a function of \mrecSquare. 

We use the $ee$ control sample to study the photon-veto performance. 
We select events with $M^2_{\rm{recoil}}<1$~GeV$^2$/$c^4$, since they come dominantly from Bhabha scattering  with a single radiated photon.
The results indicate that the photon-veto inefficiency in the backward barrel ECL is larger than that estimated in simulation. 
This study was performed after observing a large data-simulation disagreement in the signal region compatible with photon-veto inefficiency.
The photon-veto inefficiencies measured with the $ee$ control sample are used to correct the expected $\mu\mu$ background.

We estimate systematic uncertainties on the signal efficiency and on the signal and background template shapes. The uncertainties on the template shapes independently affect each of the  bins contained within the templates.

Uncertainties in selection efficiencies due to data-simulation mismodeling are studied by comparing data and simulation in the $\mu\mu\gamma$ and $e\mu$ control samples in three 
\mrecSquare\ ranges: [$-0.5,9$], [$9,36$], [$36,81$] GeV$^2$/$c^4$.
The two control samples provide complementary coverage of the \mrecSquare\ range, with $\mu\mu\gamma$ addressing the lower region and $e\mu$ covering the higher.
Systematic uncertainties due to data-simulation mismodeling in the trigger, luminosity, tracking efficiency, muon identification, background cross sections, and effect of the selections  are collectively evaluated through data-simulation comparison before the application of the Punzi-net.
Systematic uncertainties due to  the Punzi-net selection-efficiency differences in data and simulation are evaluated  by studying  
its efficiencies,  as they are indicators of the performances  
for the signal-like background component.
The differences from unity of the data-to-simulation ratios of event yields before the Punzi-net application and of the Punzi-net efficiencies in the three \mrecSquare\ ranges 
are summed in quadrature and found to be 2.7\%, 6.5\%, and 8.3\%, respectively. These differences are assigned as systematic uncertainties on the signal efficiency.
 
The recoil mass resolution is studied using the $\mu\mu\gamma$ sample.
The width of the \mrecSquare\ distribution is 8\% larger in data than in simulation. This   translates to a systematic uncertainty of 10\% on the signal template shape.  

Systematic uncertainties due to  
background shapes  are evaluated using the $\mu\mu\gamma$ and $e\mu$  samples. 
We compute the standard deviation of the bin-by-bin data-to-simulation ratios of the number of events for each search window. To be conservative, we assign twice the largest of these standard deviations in each of the three \mrecSquare\ ranges as an uncertainty for the shape in the respective \mrecSquare\ ranges. 
We use the $\mu\mu\gamma$ control sample for \mrecSquare\ up to 56\gevsq\ and the $e\mu$ control sample above. The resulting uncertainties are 3.2\%, 8.6\%, and 25\% in the three \mrecSquare\  ranges.  

Uncertainties on the background template shape from the photon-veto inefficiency are studied using the $ee$ control sample and are on average  34\% for \mrecSquare $<$ 1 \gevsq, decreasing to 5\% above 1\gevsq.
We assign a systematic uncertainty of 1\% to the measured  integrated luminosity~\cite{lumi}.

\begin{figure}[!htb]
  \centering
  \includegraphics[width=0.9\columnwidth]{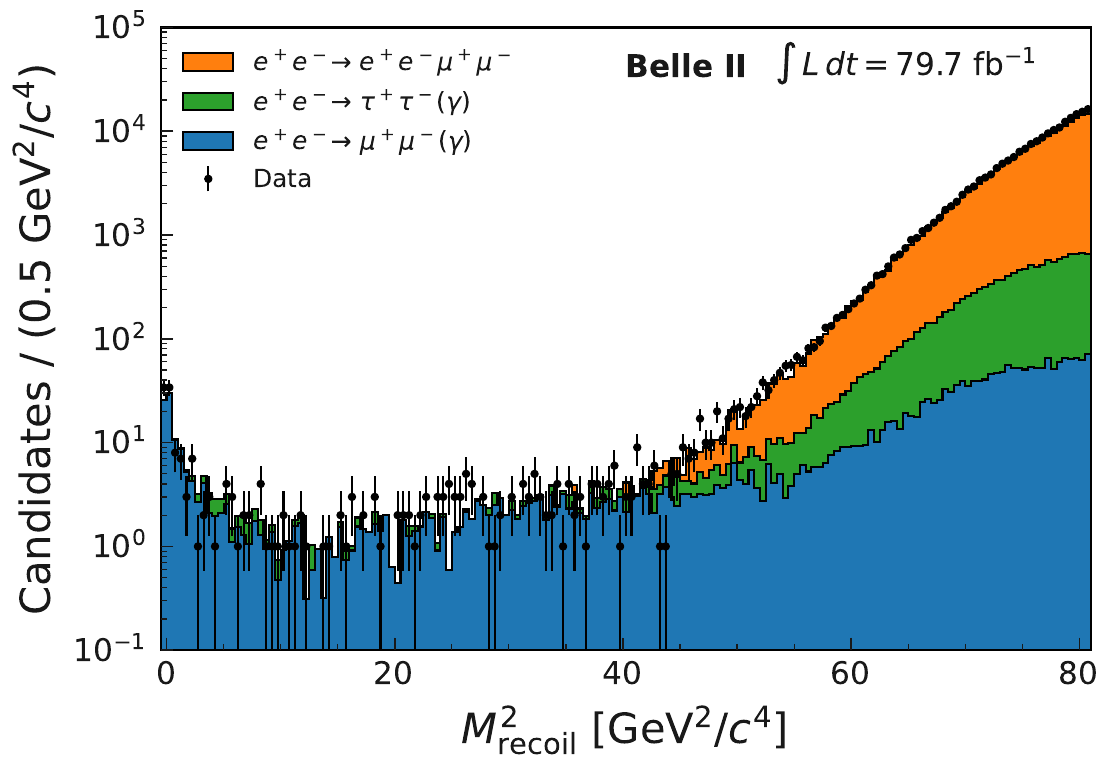}
  \caption{Squared recoil mass spectrum of the \mumu sample, compared with the stacked contributions from the various simulated background samples normalized (for illustrative purposes) to the integrated luminosity.}
  \label{fig:LP-mumu_after}
\end{figure}

The observed and expected \mrecSquare\ distributions 
are shown in Fig.~\ref{fig:LP-mumu_after}. We find no significant excess of data above the expected background. 
The $\chi^2$ value describing the goodness of the two-dimensional fit is acceptable for each test \zprime~mass with the largest incompatibility corresponding to a p value of 0.05.
The largest 
local significance is 2.8$\sigma$ for $M_{Z^\prime}=$ 2.352 GeV/$c^2$.
The global significance of this excess after correcting for the look-elsewhere effect~\cite{Gross:2010qma} is 0.7$\sigma$.

\begin{figure}[htb]
  \centering
  \includegraphics[width=0.99\columnwidth]{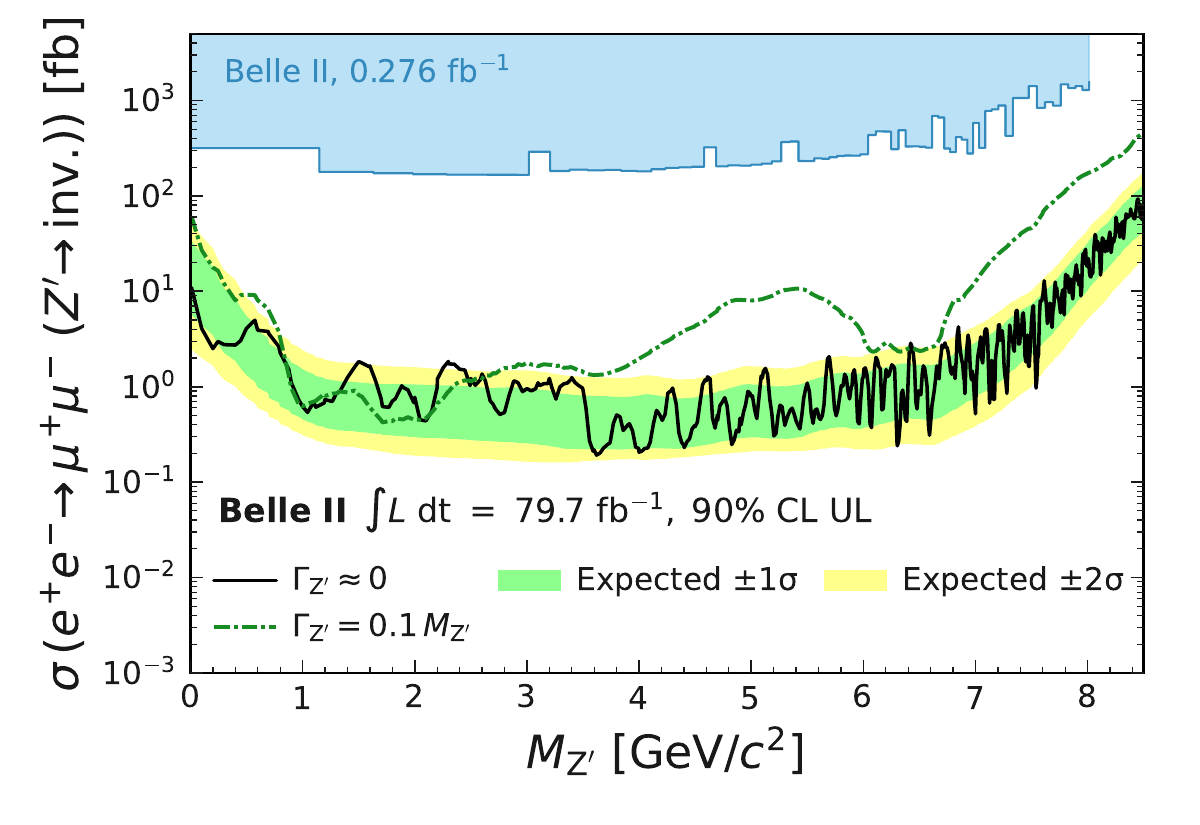}
  \caption{Observed 90$\%$ C.L. upper limits on the cross section $\sigma({e^+e^- \to \mu^+\mu^-\rm{Z}^\prime,\,\,\rm{Z}^\prime\to\rm{invisible}})$ as functions of the \zprime\ mass for the cases of negligible $\Gamma_{Z'}$ and for  $\Gamma_{\rm{Z}'}=0.1 M_{\rm{Z}'}$.
  Also shown are previous limits from Belle~II~\cite{PhysRevLett.124.141801}.}
  \label{fig:cross_section_intervals}
\end{figure}

The 90\% C.L. upper limits on the cross section for the process $e^+e^- \to \mu^+\mu^- Z^\prime$ with \zprime\ invisible, 
$\sigma({e^+e^- \to \mu^+\mu^-Z^\prime,\,\,Z^\prime\to\rm{invisible}})$ $=$ $\sigma({e^+e^- \to \mu^+\mu^-Z^\prime) \times \mathcal{B} (Z^\prime\to\rm{invisible}})$,
are shown in Fig.~\ref{fig:cross_section_intervals} as functions of $M_{Z^\prime}$, along with the  1$\sigma$ and 2$\sigma$ bands of expected limits (the median limits from background-only simulated samples).
We set upper limits as small as 0.2 fb.
In addition, we show upper limits for the benchmark scenario in which we assume non-negligible $\Gamma_{Z^\prime}$. Our upper limits are dominated by statistical uncertainties for $M_{Z^\prime}$ < 6 \gevcc, where systematic uncertainties degrade them by less than 5\%.  Above 6 \gevcc, upper limits are dominated by systematic uncertainties (mainly due to background shapes), degrading them by about 40\%.

Cross section results are translated into 90\% CL upper limits on the coupling  $g^{\prime}$. 
In both fully invisible and vanilla 
models, we  focus on the direct-search results and do not show constraints obtained from reanalyses of data from neutrino experiments~\cite{PhysRevLett.113.091801, PhysRevLett.107.141302, Kamada2018}.

\begin{figure}[!htb]
  \centering
  \includegraphics[width=0.99\columnwidth]{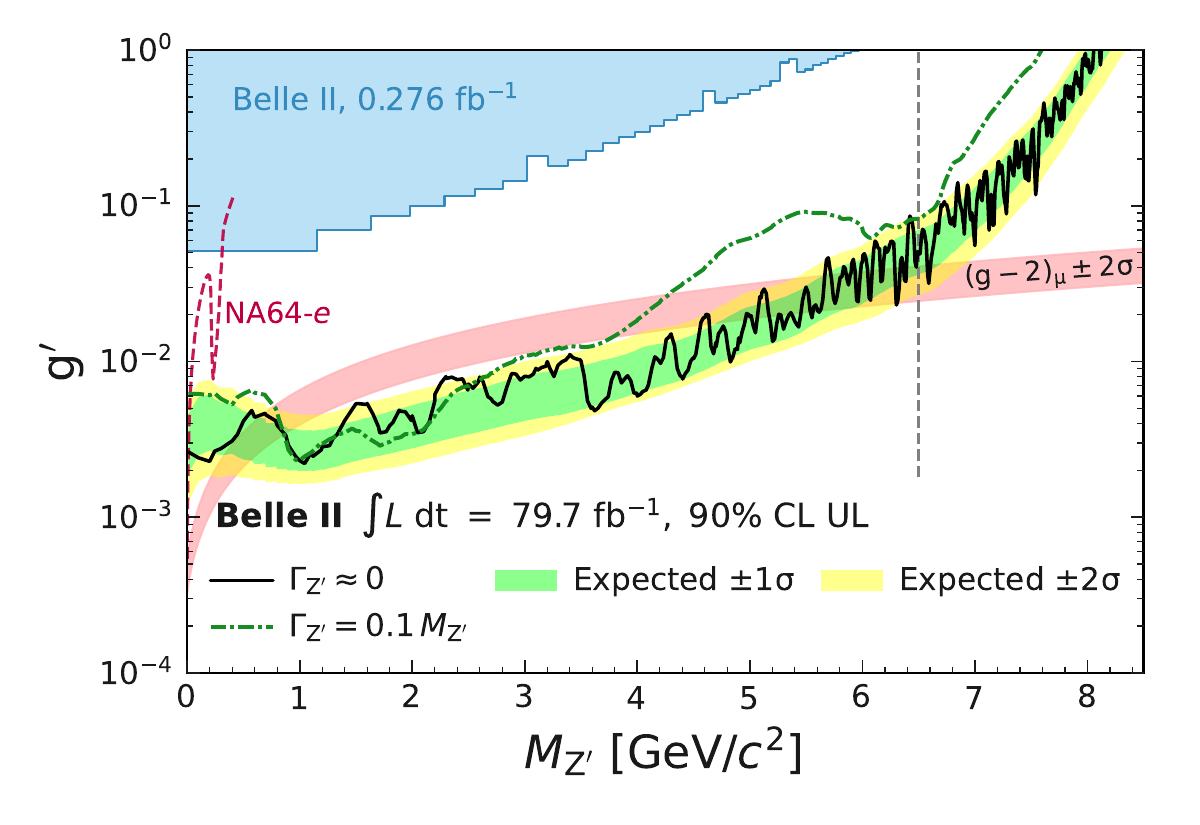}
  \caption{Observed 90$\%$ C.L. upper limits on the coupling   $g^{\prime}$ for the  fully invisible \lmultau\ model as functions of the \zprime\ mass for the cases of negligible $\Gamma_{\rm{Z}'}$ and for  $\Gamma_{\rm{Z}'}=0.1 M_{Z'}$.
    Also shown are previous limits from NA64-$e$~\cite{Andreev:2022txy} and Belle~II~\cite{PhysRevLett.124.141801} searches.
  The red band shows the region that explains the measured value of the muon anomalous magnetic moment $(g-2)_{\mu} \pm 2\sigma$~\cite{PhysRevLett.126.141801}. The vertical dashed line indicates the limit beyond which the hypothesis $\cal{B}\it(\rm{Z}^{\prime} \to \chi \bar{\chi}) \approx$~1 is not respected in the negligible $\Gamma_{\rm{Z}'}$ case.}
  \label{fig:gprime_intervals_invisible}
\end{figure}

Figure~\ref{fig:gprime_intervals_invisible}  presents limits in the  fully invisible \lmultau\ model for the cases of negligible and non-negligible  $\Gamma_{\rm{Z}^\prime}$.
For the case of negligible $\Gamma_{\rm{Z}^\prime}$, these constraints hold for \zprimemass~$\lesssim$~6.5~\gevcc. 
Above this mass, there is no value of \alphad\ that produces both a negligible width and $\cal{B}\it(\rm{Z}^{\prime} \to \chi \bar{\chi}) \approx \rm1$, given the values of $g^{\prime}$ being probed. Numerical values in Fig.~\ref{fig:gprime_intervals_invisible} can still be used, but need to be rescaled by 
$1 / \sqrt{ \cal{B}\it(\rm{Z}^{\prime} \to \chi \bar{\chi})}$, which depends on \alphad.
We also show limits from NA64-$e$~\cite{Andreev:2022txy} and the previous Belle~II search~\cite{PhysRevLett.124.141801}.
Our results are world-leading for direct searches of \zprime\ with masses above 11.5~\mevcc. They are the first direct-search results to exclude at 90\% C.L. the fully invisible-\zprime\ model as an explanation of the $(g-2)_\mu$ anomaly for $0.8<M_{\rm{Z}^\prime}<5.0$ \gevcc.

\begin{figure}[!htb]
  \centering
  \includegraphics[width=0.99\columnwidth]{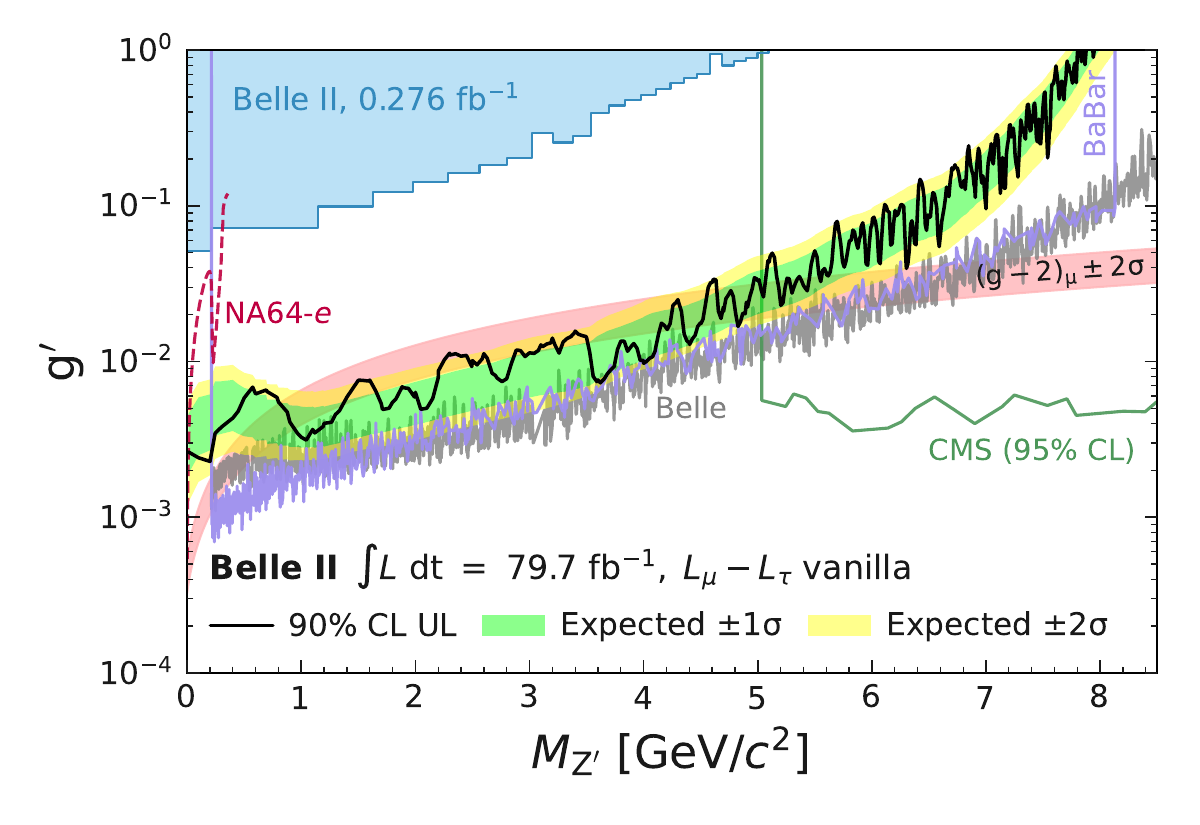}
  \caption{Observed 90$\%$ C.L. upper limits on the coupling $g^{\prime}$ for the  vanilla \lmultau\ model as functions of the \zprime\ mass.
  Also shown are previous limits from  Belle~II~\cite{PhysRevLett.124.141801} and NA64-$e$~\cite{Andreev:2022txy} searches for invisible \zprime\ decays, and from Belle~\cite{Belle:2021feg}, \textit{BaBar}~\cite{TheBABAR:2016rlg}, and CMS~\cite{cms}  searches for \zprime\ decays to muons (at 95\% C.L.). The red band shows the region that  explains the muon anomalous magnetic moment $(g-2)_{\mu} \pm 2\sigma$~\cite{PhysRevLett.126.141801}.}
  \label{fig:gprime_intervals_lmu_ltau}
\end{figure}

Figure~\ref{fig:gprime_intervals_lmu_ltau} presents limits in the  vanilla \lmultau\ model. Our results are world leading for direct searches of \zprime\  in the mass  range 11.5 to 211 \mevcc. More stringent limits are from NA64-$e$~\cite{PhysRevLett.124.141801} below 11 \mevcc and from Belle~\cite{Belle:2021feg},  \textit{BABAR} \cite{TheBABAR:2016rlg}, and CMS~\cite{cms} searches for Z$^\prime\to\mu^+\mu^-$  above 211 \mevcc.

Additional plots, including indirect constraints from neutrino experiments and  detailed numerical results, are provided in the Supplemental Material~\cite{supplemental}.

In summary, we search for an invisibly decaying \zprime\ boson in the process $e^+e^- \to \mu^+\mu^-Z^{\prime}$ using data corresponding to 79.7\invfb\ collected by Belle~II at SuperKEKB in 2019--2020.
We find no significant excess above the expected background and set 90\% C.L. upper limits on the coupling  $g^{\prime}$  ranging from $3 \times 10^{-3}$ at low \zprime\ masses to 1 for a mass of 8~\gevcc.
These are world-leading direct-search results for \zprime\ masses above 11.5~\mevcc in the  fully invisible \lmultau\ model and for masses in the range 11.5 to 211 \mevcc\ 
in the  vanilla \lmultau\ model. 
These limits are the first direct-search results excluding a fully invisible-\zprime-boson model as an explanation of the $(g-2)_\mu$ anomaly for $0.8<M_{\rm{Z}^\prime}<5.0$ \gevcc.


We thank Andreas Crivellin for useful discussions during the preparation of this manuscript.
This work, based on data collected using the Belle II detector, which was built and commissioned prior to March 2019, was supported by
the Science Committee of the Republic of Armenia Grant No.~20TTCG-1C010;
Australian Research Council and Research Grants
No.~DE220100462,
No.~DP180102629,
No.~DP170102389,
No.~DP170102204,
No.~DP150103061,
No.~FT130100303,
No.~FT130100018,
and
No.~FT120100745;
Austrian Federal Ministry of Education, Science and Research,
Austrian Science Fund
No.~P~31361-N36
and
No.~J4625-N,
and
Horizon 2020 ERC Starting Grant No.~947006 ``InterLeptons'';
Natural Sciences and Engineering Research Council of Canada, Compute Canada and CANARIE;
Chinese Academy of Sciences and Research Grant No.~QYZDJ-SSW-SLH011,
National Natural Science Foundation of China and Research Grants
No.~11521505,
No.~11575017,
No.~11675166,
No.~11761141009,
No.~11705209,
and
No.~11975076,
LiaoNing Revitalization Talents Program under Contract No.~XLYC1807135,
Shanghai Pujiang Program under Grant No.~18PJ1401000,
and the CAS Center for Excellence in Particle Physics (CCEPP);
the Ministry of Education, Youth, and Sports of the Czech Republic under Contract No.~LTT17020 and
Charles University Grant No.~SVV 260448 and
the Czech Science Foundation Grant No.~22-18469S;
European Research Council, Seventh Framework PIEF-GA-2013-622527,
Horizon 2020 ERC-Advanced Grants No.~267104 and No.~884719,
Horizon 2020 ERC-Consolidator Grant No.~819127,
Horizon 2020 Marie Sklodowska-Curie Grant Agreement No.~700525 "NIOBE"
and
No.~101026516,
and
Horizon 2020 Marie Sklodowska-Curie RISE project JENNIFER2 Grant Agreement No.~822070 (European grants);
L'Institut National de Physique Nucl\'{e}aire et de Physique des Particules (IN2P3) du CNRS (France);
BMBF, DFG, HGF, MPG, and AvH Foundation (Germany);
Department of Atomic Energy under Project Identification No.~RTI 4002 and Department of Science and Technology (India);
Israel Science Foundation Grant No.~2476/17,
U.S.-Israel Binational Science Foundation Grant No.~2016113, and
Israel Ministry of Science Grant No.~3-16543;
Istituto Nazionale di Fisica Nucleare and the Research grants BELLE2;
Japan Society for the Promotion of Science, Grant-in-Aid for Scientific Research Grants
No.~16H03968,
No.~16H03993,
No.~16H06492,
No.~16K05323,
No.~17H01133,
No.~17H05405,
No.~18K03621,
No.~18H03710,
No.~18H05226,
No.~19H00682, 
No.~22H00144,
No.~26220706,
and
No.~26400255,
the National Institute of Informatics, and Science Information NETwork 5 (SINET5), 
and
the Ministry of Education, Culture, Sports, Science, and Technology (MEXT) of Japan;  
National Research Foundation (NRF) of Korea Grants
No.~2016R1\-D1A1B\-02012900,
No.~2018R1\-A2B\-3003643,
No.~2018R1\-A6A1A\-06024970,
No.~2018R1\-D1A1B\-07047294,
No.~2019R1\-I1A3A\-01058933,
No.~2021R1\-A4A\-2001897,
No.~2022R1\-A2C\-1003993,
and
No.~RS-2022-00197659,
Radiation Science Research Institute,
Foreign Large-size Research Facility Application Supporting project,
the Global Science Experimental Data Hub Center of the Korea Institute of Science and Technology Information
and
KREONET/GLORIAD;
Universiti Malaya RU grant, Akademi Sains Malaysia, and Ministry of Education Malaysia;
Frontiers of Science Program Contracts
No.~FOINS-296,
No.~CB-221329,
No.~CB-236394,
No.~CB-254409,
and
No.~CB-180023, and No.~SEP-CINVESTAV Research Grant No.~237 (Mexico);
the Polish Ministry of Science and Higher Education and the National Science Center;
the Ministry of Science and Higher Education of the Russian Federation,
Agreement No.~14.W03.31.0026, and
the HSE University Basic Research Program, Moscow;
University of Tabuk Research Grants
No.~S-0256-1438 and No.~S-0280-1439 (Saudi Arabia);
Slovenian Research Agency and Research Grants
No.~J1-9124
and
No.~P1-0135;
Agencia Estatal de Investigacion, Spain
Grant No.~RYC2020-029875-I
and
Generalitat Valenciana, Spain
Grant No.~CIDEGENT/2018/020
Ministry of Science and Technology and Research Grants
No.~MOST106-2112-M-002-005-MY3
and
No.~MOST107-2119-M-002-035-MY3,
and the Ministry of Education (Taiwan);
Thailand Center of Excellence in Physics;
TUBITAK ULAKBIM (Turkey);
National Research Foundation of Ukraine, Project No.~2020.02/0257,
and
Ministry of Education and Science of Ukraine;
the U.S. National Science Foundation and Research Grants
No.~PHY-1913789 
and
No.~PHY-2111604, 
and the U.S. Department of Energy and Research Awards
No.~DE-AC06-76RLO1830, 
No.~DE-SC0007983, 
No.~DE-SC0009824, 
No.~DE-SC0009973, 
No.~DE-SC0010007, 
No.~DE-SC0010073, 
No.~DE-SC0010118, 
No.~DE-SC0010504, 
No.~DE-SC0011784, 
No.~DE-SC0012704, 
No.~DE-SC0019230, 
No.~DE-SC0021274 and 
No.~DE-SC0022350; 
and
the Vietnam Academy of Science and Technology (VAST) under Grant No.~DL0000.05/21-23.

These acknowledgements are not to be interpreted as an endorsement of any statement made
by any of our institutes, funding agencies, governments, or their representatives. 

We thank the SuperKEKB team for delivering high-luminosity collisions;
the KEK cryogenics group for the efficient operation of the detector solenoid magnet;
the KEK computer group and the NII for on site computing support and SINET6 network support;
the raw-data centers at BNL, DESY, GridKa, IN2P3, INFN, and the University of Victoria for off site computing support.

\bibliography{bibliography.bib}

\begin{thebibliography}{50}
\expandafter\ifx\csname natexlab\endcsname\relax\def\natexlab#1{#1}\fi
\expandafter\ifx\csname bibnamefont\endcsname\relax
  \def\bibnamefont#1{#1}\fi
\expandafter\ifx\csname bibfnamefont\endcsname\relax
  \def\bibfnamefont#1{#1}\fi
\expandafter\ifx\csname citenamefont\endcsname\relax
  \def\citenamefont#1{#1}\fi
\expandafter\ifx\csname url\endcsname\relax
  \def\url#1{\texttt{#1}}\fi
\expandafter\ifx\csname urlprefix\endcsname\relax\def\urlprefix{URL }\fi
\providecommand{\bibinfo}[2]{#2}
\providecommand{\eprint}[2][]{\url{#2}}

\bibitem[{\citenamefont{Bennett et~al.}(2006)}]{PhysRevD.73.072003}
\bibinfo{author}{\bibfnamefont{G.~W.} \bibnamefont{Bennett}}
  \bibnamefont{et~al.} (\bibinfo{collaboration}{Muon $g\ensuremath{-}2$
  Collaboration}), \bibinfo{journal}{Phys. Rev. D}
  \textbf{\bibinfo{volume}{73}}, \bibinfo{pages}{072003}
  (\bibinfo{year}{2006}).

\bibitem[{\citenamefont{Abi et~al.}(2021)}]{PhysRevLett.126.141801}
\bibinfo{author}{\bibfnamefont{B.}~\bibnamefont{Abi}} \bibnamefont{et~al.}
  (\bibinfo{collaboration}{Muon $g\ensuremath{-}2$ Collaboration}),
  \bibinfo{journal}{Phys. Rev. Lett.} \textbf{\bibinfo{volume}{126}},
  \bibinfo{pages}{141801} (\bibinfo{year}{2021}).

\bibitem[{\citenamefont{Bertone et~al.}(2005)\citenamefont{Bertone, Hooper, and
  Silk}}]{BERTONE2005279}
\bibinfo{author}{\bibfnamefont{G.}~\bibnamefont{Bertone}},
  \bibinfo{author}{\bibfnamefont{D.}~\bibnamefont{Hooper}}, \bibnamefont{and}
  \bibinfo{author}{\bibfnamefont{J.}~\bibnamefont{Silk}},
  \bibinfo{journal}{Phys. Rep.} \textbf{\bibinfo{volume}{405}},
  \bibinfo{pages}{279} (\bibinfo{year}{2005}).

\bibitem[{\citenamefont{He et~al.}(1991)\citenamefont{He, Joshi, Lew, and
  Volkas}}]{PhysRevD.43.R22}
\bibinfo{author}{\bibfnamefont{X.~G.} \bibnamefont{He}},
  \bibinfo{author}{\bibfnamefont{G.~C.} \bibnamefont{Joshi}},
  \bibinfo{author}{\bibfnamefont{H.}~\bibnamefont{Lew}}, \bibnamefont{and}
  \bibinfo{author}{\bibfnamefont{R.~R.} \bibnamefont{Volkas}},
  \bibinfo{journal}{Phys. Rev. D} \textbf{\bibinfo{volume}{43}},
  \bibinfo{pages}{R22} (\bibinfo{year}{1991}).

\bibitem[{\citenamefont{Shuve and Yavin}(2014)}]{Shuve:2014doa}
\bibinfo{author}{\bibfnamefont{B.}~\bibnamefont{Shuve}} \bibnamefont{and}
  \bibinfo{author}{\bibfnamefont{I.}~\bibnamefont{Yavin}},
  \bibinfo{journal}{Phys. Rev. D} \textbf{\bibinfo{volume}{89}},
  \bibinfo{pages}{113004} (\bibinfo{year}{2014}).

\bibitem[{\citenamefont{Altmannshofer et~al.}(2016)\citenamefont{Altmannshofer,
  Gori, Profumo, and Queiroz}}]{Altmannshofer:2016jzy}
\bibinfo{author}{\bibfnamefont{W.}~\bibnamefont{Altmannshofer}},
  \bibinfo{author}{\bibfnamefont{S.}~\bibnamefont{Gori}},
  \bibinfo{author}{\bibfnamefont{S.}~\bibnamefont{Profumo}}, \bibnamefont{and}
  \bibinfo{author}{\bibfnamefont{F.~S.} \bibnamefont{Queiroz}},
  \bibinfo{journal}{J. High Energy Phys.} \textbf{\bibinfo{volume}{12}},
  \bibinfo{pages}{106} (\bibinfo{year}{2016}).

\bibitem[{\citenamefont{Altmannshofer et~al.}(2014)\citenamefont{Altmannshofer,
  Gori, Pospelov, and Yavin}}]{PhysRevLett.113.091801}
\bibinfo{author}{\bibfnamefont{W.}~\bibnamefont{Altmannshofer}},
  \bibinfo{author}{\bibfnamefont{S.}~\bibnamefont{Gori}},
  \bibinfo{author}{\bibfnamefont{M.}~\bibnamefont{Pospelov}}, \bibnamefont{and}
  \bibinfo{author}{\bibfnamefont{I.}~\bibnamefont{Yavin}},
  \bibinfo{journal}{Phys. Rev. Lett.} \textbf{\bibinfo{volume}{113}},
  \bibinfo{pages}{091801} (\bibinfo{year}{2014}).

\bibitem[{\citenamefont{Aaij et~al.}(2022)}]{LHCb:2021trn}
\bibinfo{author}{\bibfnamefont{R.}~\bibnamefont{Aaij}} \bibnamefont{et~al.}
  (\bibinfo{collaboration}{LHCb Collaboration}), \bibinfo{journal}{Nature
  Phys.} \textbf{\bibinfo{volume}{18}}, \bibinfo{pages}{277}
  (\bibinfo{year}{2022}).

\bibitem[{\citenamefont{Lees et~al.}(2013)}]{BaBar:2013mob}
\bibinfo{author}{\bibfnamefont{J.~P.} \bibnamefont{Lees}} \bibnamefont{et~al.}
  (\bibinfo{collaboration}{BaBar Collaboration}), \bibinfo{journal}{Phys. Rev.
  D} \textbf{\bibinfo{volume}{88}}, \bibinfo{pages}{072012}
  (\bibinfo{year}{2013}), \eprint{1303.0571}.

\bibitem[{\citenamefont{Huschle et~al.}(2015)}]{Belle:2015qfa}
\bibinfo{author}{\bibfnamefont{M.}~\bibnamefont{Huschle}} \bibnamefont{et~al.}
  (\bibinfo{collaboration}{Belle Collaboration}), \bibinfo{journal}{Phys. Rev.
  D} \textbf{\bibinfo{volume}{92}}, \bibinfo{pages}{072014}
  (\bibinfo{year}{2015}), \eprint{1507.03233}.

\bibitem[{\citenamefont{Aaij et~al.}(2015)}]{LHCb:2015gmp}
\bibinfo{author}{\bibfnamefont{R.}~\bibnamefont{Aaij}} \bibnamefont{et~al.}
  (\bibinfo{collaboration}{LHCb Collaboration}), \bibinfo{journal}{Phys. Rev.
  Lett.} \textbf{\bibinfo{volume}{115}}, \bibinfo{pages}{111803}
  (\bibinfo{year}{2015}), \bibinfo{note}{115, 159901(E) (2015)}.

\bibitem[{\citenamefont{Hirose et~al.}(2018)}]{Belle:2017ilt}
\bibinfo{author}{\bibfnamefont{S.}~\bibnamefont{Hirose}} \bibnamefont{et~al.}
  (\bibinfo{collaboration}{Belle Collaboration}), \bibinfo{journal}{Phys. Rev.
  D} \textbf{\bibinfo{volume}{97}}, \bibinfo{pages}{012004}
  (\bibinfo{year}{2018}), \eprint{1709.00129}.

\bibitem[{\citenamefont{Aaij et~al.}(2018)}]{LHCb:2017rln}
\bibinfo{author}{\bibfnamefont{R.}~\bibnamefont{Aaij}} \bibnamefont{et~al.}
  (\bibinfo{collaboration}{LHCb Collaboration}), \bibinfo{journal}{Phys. Rev.
  D} \textbf{\bibinfo{volume}{97}}, \bibinfo{pages}{072013}
  (\bibinfo{year}{2018}), \eprint{1711.02505}.

\bibitem[{\citenamefont{Caria et~al.}(2020)}]{Belle:2019rba}
\bibinfo{author}{\bibfnamefont{G.}~\bibnamefont{Caria}} \bibnamefont{et~al.}
  (\bibinfo{collaboration}{Belle Collaboration}), \bibinfo{journal}{Phys. Rev.
  Lett.} \textbf{\bibinfo{volume}{124}}, \bibinfo{pages}{161803}
  (\bibinfo{year}{2020}), \eprint{1910.05864}.

\bibitem[{\citenamefont{Crivellin et~al.}(2022)\citenamefont{Crivellin,
  Manzari, Altmannshofer, Inguglia, Feichtinger, and
  M.~Camalich}}]{Crivellin:2022obd}
\bibinfo{author}{\bibfnamefont{A.}~\bibnamefont{Crivellin}},
  \bibinfo{author}{\bibfnamefont{C.~A.} \bibnamefont{Manzari}},
  \bibinfo{author}{\bibfnamefont{W.}~\bibnamefont{Altmannshofer}},
  \bibinfo{author}{\bibfnamefont{G.}~\bibnamefont{Inguglia}},
  \bibinfo{author}{\bibfnamefont{P.}~\bibnamefont{Feichtinger}},
  \bibnamefont{and}
  \bibinfo{author}{\bibfnamefont{J.}~\bibnamefont{M.~Camalich}},
  \bibinfo{journal}{Phys. Rev. D} \textbf{\bibinfo{volume}{106}},
  \bibinfo{pages}{L031703} (\bibinfo{year}{2022}).

\bibitem[{\citenamefont{Sala and Straub}(2017)}]{SALA2017205}
\bibinfo{author}{\bibfnamefont{F.}~\bibnamefont{Sala}} \bibnamefont{and}
  \bibinfo{author}{\bibfnamefont{D.~M.} \bibnamefont{Straub}},
  \bibinfo{journal}{Phys. Lett. B} \textbf{\bibinfo{volume}{774}},
  \bibinfo{pages}{205} (\bibinfo{year}{2017}).

\bibitem[{\citenamefont{Chen and Nomura}(2018)}]{Chen:2017usq}
\bibinfo{author}{\bibfnamefont{C.-H.} \bibnamefont{Chen}} \bibnamefont{and}
  \bibinfo{author}{\bibfnamefont{T.}~\bibnamefont{Nomura}},
  \bibinfo{journal}{Phys. Lett. B} \textbf{\bibinfo{volume}{777}},
  \bibinfo{pages}{420} (\bibinfo{year}{2018}), \eprint{1707.03249}.

\bibitem[{\citenamefont{Greljo et~al.}(2021)\citenamefont{Greljo, Stangl, and
  Thomsen}}]{Greljo:2021xmg}
\bibinfo{author}{\bibfnamefont{A.}~\bibnamefont{Greljo}},
  \bibinfo{author}{\bibfnamefont{P.}~\bibnamefont{Stangl}}, \bibnamefont{and}
  \bibinfo{author}{\bibfnamefont{A.~E.} \bibnamefont{Thomsen}},
  \bibinfo{journal}{Phys. Lett. B} \textbf{\bibinfo{volume}{820}},
  \bibinfo{pages}{136554} (\bibinfo{year}{2021}), \eprint{2103.13991}.

\bibitem[{\citenamefont{Araki et~al.}(2017)\citenamefont{Araki, Hoshino, Ota,
  Sato, and Shimomura}}]{PhysRevD.95.055006}
\bibinfo{author}{\bibfnamefont{T.}~\bibnamefont{Araki}},
  \bibinfo{author}{\bibfnamefont{S.}~\bibnamefont{Hoshino}},
  \bibinfo{author}{\bibfnamefont{T.}~\bibnamefont{Ota}},
  \bibinfo{author}{\bibfnamefont{J.}~\bibnamefont{Sato}}, \bibnamefont{and}
  \bibinfo{author}{\bibfnamefont{T.}~\bibnamefont{Shimomura}},
  \bibinfo{journal}{Phys. Rev. D} \textbf{\bibinfo{volume}{95}},
  \bibinfo{pages}{055006} (\bibinfo{year}{2017}).

\bibitem[{\citenamefont{Bauer et~al.}(2018)\citenamefont{Bauer, Foldenauer, and
  Jaeckel}}]{hunting}
\bibinfo{author}{\bibfnamefont{M.}~\bibnamefont{Bauer}},
  \bibinfo{author}{\bibfnamefont{P.}~\bibnamefont{Foldenauer}},
  \bibnamefont{and} \bibinfo{author}{\bibfnamefont{J.}~\bibnamefont{Jaeckel}},
  \bibinfo{journal}{J. High Energy Phys.} \textbf{\bibinfo{volume}{07}},
  \bibinfo{pages}{094} (\bibinfo{year}{2018}).

\bibitem[{\citenamefont{Lees et~al.}(2016)}]{TheBABAR:2016rlg}
\bibinfo{author}{\bibfnamefont{J.~P.} \bibnamefont{Lees}} \bibnamefont{et~al.}
  (\bibinfo{collaboration}{\textit{BABAR} Collaboration}),
  \bibinfo{journal}{Phys. Rev. D} \textbf{\bibinfo{volume}{94}},
  \bibinfo{pages}{011102} (\bibinfo{year}{2016}).

\bibitem[{\citenamefont{Czank et~al.}(2022)}]{Belle:2021feg}
\bibinfo{author}{\bibfnamefont{T.}~\bibnamefont{Czank}} \bibnamefont{et~al.}
  (\bibinfo{collaboration}{Belle Collaboration}), \bibinfo{journal}{Phys. Rev.
  D} \textbf{\bibinfo{volume}{106}}, \bibinfo{pages}{012003}
  (\bibinfo{year}{2022}).

\bibitem[{\citenamefont{Sirunyan et~al.}(2019)}]{cms}
\bibinfo{author}{\bibfnamefont{A.~M.} \bibnamefont{Sirunyan}}
  \bibnamefont{et~al.} (\bibinfo{collaboration}{CMS Collaboration}),
  \bibinfo{journal}{Phys. Lett. B} \textbf{\bibinfo{volume}{792}},
  \bibinfo{pages}{345} (\bibinfo{year}{2019}).

\bibitem[{\citenamefont{{ATLAS Collaboration}}(2023)}]{ATLAS:2023vxg}
\bibinfo{author}{\bibnamefont{{ATLAS Collaboration}}} (\bibinfo{year}{2023}),
  \eprint{arXiv:2301.09342}.

\bibitem[{\citenamefont{Andreev et~al.}(2022)}]{Andreev:2022txy}
\bibinfo{author}{\bibfnamefont{Y.~M.} \bibnamefont{Andreev}}
  \bibnamefont{et~al.} (\bibinfo{collaboration}{NA64 Collaboration}),
  \bibinfo{journal}{Phys. Rev. D} \textbf{\bibinfo{volume}{106}},
  \bibinfo{pages}{032015} (\bibinfo{year}{2022}).

\bibitem[{\citenamefont{Adachi et~al.}(2020)}]{PhysRevLett.124.141801}
\bibinfo{author}{\bibfnamefont{I.}~\bibnamefont{Adachi}} \bibnamefont{et~al.}
  (\bibinfo{collaboration}{Belle II Collaboration}), \bibinfo{journal}{Phys.
  Rev. Lett.} \textbf{\bibinfo{volume}{124}}, \bibinfo{pages}{141801}
  (\bibinfo{year}{2020}).

\bibitem[{\citenamefont{Abudin{\'{e}}n et~al.}(2020)}]{lumi}
\bibinfo{author}{\bibfnamefont{F.}~\bibnamefont{Abudin{\'{e}}n}}
  \bibnamefont{et~al.} (\bibinfo{collaboration}{Belle II Collaboration}),
  \bibinfo{journal}{Chin. Phys. C} \textbf{\bibinfo{volume}{44}},
  \bibinfo{pages}{021001} (\bibinfo{year}{2020}).

\bibitem[{\citenamefont{Abe et~al.}(2010)}]{Abe:2010sj}
\bibinfo{author}{\bibfnamefont{T.}~\bibnamefont{Abe}} \bibnamefont{et~al.}
  (\bibinfo{collaboration}{Belle II Collaboration}) (\bibinfo{year}{2010}),
  \eprint{arXiv:1011.0352}.

\bibitem[{\citenamefont{Akai et~al.}(2018)\citenamefont{Akai, Furukawa, and
  Koiso}}]{superkekb}
\bibinfo{author}{\bibfnamefont{K.}~\bibnamefont{Akai}},
  \bibinfo{author}{\bibfnamefont{K.}~\bibnamefont{Furukawa}}, \bibnamefont{and}
  \bibinfo{author}{\bibfnamefont{H.}~\bibnamefont{Koiso}}
  (\bibinfo{collaboration}{SuperKEKB Accelerator Team}),
  \bibinfo{journal}{Nucl. Instrum. Methods Phys. Res. A}
  \textbf{\bibinfo{volume}{907}}, \bibinfo{pages}{188} (\bibinfo{year}{2018}).

\bibitem[{\citenamefont{Kou et~al.}(2019)}]{ref:b2tip}
\bibinfo{author}{\bibfnamefont{E.}~\bibnamefont{Kou}} \bibnamefont{et~al.},
  \bibinfo{journal}{Prog. Theor. Exp. Phys.} \textbf{\bibinfo{volume}{2019}},
  \bibinfo{pages}{123C01} (\bibinfo{year}{2019}).

\bibitem[{\citenamefont{Alwall et~al.}(2014)\citenamefont{Alwall, Frederix,
  Frixione, Hirschi, Maltoni, Mattelaer, Shao, Stelzer, Torrielli, and
  Zaro}}]{Alwall2014}
\bibinfo{author}{\bibfnamefont{J.}~\bibnamefont{Alwall}},
  \bibinfo{author}{\bibfnamefont{R.}~\bibnamefont{Frederix}},
  \bibinfo{author}{\bibfnamefont{S.}~\bibnamefont{Frixione}},
  \bibinfo{author}{\bibfnamefont{V.}~\bibnamefont{Hirschi}},
  \bibinfo{author}{\bibfnamefont{F.}~\bibnamefont{Maltoni}},
  \bibinfo{author}{\bibfnamefont{O.}~\bibnamefont{Mattelaer}},
  \bibinfo{author}{\bibfnamefont{H.~S.} \bibnamefont{Shao}},
  \bibinfo{author}{\bibfnamefont{T.}~\bibnamefont{Stelzer}},
  \bibinfo{author}{\bibfnamefont{P.}~\bibnamefont{Torrielli}},
  \bibnamefont{and} \bibinfo{author}{\bibfnamefont{M.}~\bibnamefont{Zaro}},
  \bibinfo{journal}{J. High Energy Phys.} \textbf{\bibinfo{volume}{07}},
  \bibinfo{pages}{079} (\bibinfo{year}{2014}).

\bibitem[{\citenamefont{Jadach et~al.}(2000)\citenamefont{Jadach, Ward, and
  W\k{a}s}}]{ref:kkmc}
\bibinfo{author}{\bibfnamefont{S.}~\bibnamefont{Jadach}},
  \bibinfo{author}{\bibfnamefont{B.~F.~L.} \bibnamefont{Ward}},
  \bibnamefont{and} \bibinfo{author}{\bibfnamefont{Z.}~\bibnamefont{W\k{a}s}},
  \bibinfo{journal}{Comput. Phys. Commun.} \textbf{\bibinfo{volume}{130}},
  \bibinfo{pages}{260} (\bibinfo{year}{2000}).

\bibitem[{\citenamefont{Davidson et~al.}(2012)\citenamefont{Davidson, Nanava,
  Przedzinski, Richter-W\k{a}s, and W\k{a}s}}]{ref:tauola}
\bibinfo{author}{\bibfnamefont{N.}~\bibnamefont{Davidson}},
  \bibinfo{author}{\bibfnamefont{G.}~\bibnamefont{Nanava}},
  \bibinfo{author}{\bibfnamefont{T.}~\bibnamefont{Przedzinski}},
  \bibinfo{author}{\bibfnamefont{E.}~\bibnamefont{Richter-W\k{a}s}},
  \bibnamefont{and} \bibinfo{author}{\bibfnamefont{Z.}~\bibnamefont{W\k{a}s}},
  \bibinfo{journal}{Comput. Phys. Commun.} \textbf{\bibinfo{volume}{183}},
  \bibinfo{pages}{821} (\bibinfo{year}{2012}).

\bibitem[{\citenamefont{Berends et~al.}(1985)\citenamefont{Berends, Daverveldt,
  and Kleiss}}]{ref:fourlepton}
\bibinfo{author}{\bibfnamefont{F.~A.} \bibnamefont{Berends}},
  \bibinfo{author}{\bibfnamefont{P.~H.} \bibnamefont{Daverveldt}},
  \bibnamefont{and} \bibinfo{author}{\bibfnamefont{R.}~\bibnamefont{Kleiss}},
  \bibinfo{journal}{Nucl. Phys. B} \textbf{\bibinfo{volume}{253}},
  \bibinfo{pages}{441} (\bibinfo{year}{1985}).

\bibitem[{\citenamefont{Czyż et~al.}(2013)\citenamefont{Czyż, Gunia, and
  Kühn}}]{ref:phokhara}
\bibinfo{author}{\bibfnamefont{H.}~\bibnamefont{Czyż}},
  \bibinfo{author}{\bibfnamefont{M.}~\bibnamefont{Gunia}}, \bibnamefont{and}
  \bibinfo{author}{\bibfnamefont{J.~H.} \bibnamefont{Kühn}},
  \bibinfo{journal}{J. High Energy Phys.} \textbf{\bibinfo{volume}{08}},
  \bibinfo{pages}{110} (\bibinfo{year}{2013}).

\bibitem[{\citenamefont{Balossini et~al.}(2008)\citenamefont{Balossini,
  Bignamini, Calame, Montagna, Nicrosini, and Piccinini}}]{ref:babayaga}
\bibinfo{author}{\bibfnamefont{G.}~\bibnamefont{Balossini}},
  \bibinfo{author}{\bibfnamefont{C.}~\bibnamefont{Bignamini}},
  \bibinfo{author}{\bibfnamefont{C.~M.~C.} \bibnamefont{Calame}},
  \bibinfo{author}{\bibfnamefont{G.}~\bibnamefont{Montagna}},
  \bibinfo{author}{\bibfnamefont{O.}~\bibnamefont{Nicrosini}},
  \bibnamefont{and}
  \bibinfo{author}{\bibfnamefont{F.}~\bibnamefont{Piccinini}},
  \bibinfo{journal}{Phys. Lett. B} \textbf{\bibinfo{volume}{663}},
  \bibinfo{pages}{209} (\bibinfo{year}{2008}).

\bibitem[{\citenamefont{Agostinelli et~al.}(2003)}]{ref:geant4}
\bibinfo{author}{\bibfnamefont{S.}~\bibnamefont{Agostinelli}}
  \bibnamefont{et~al.} (\bibinfo{collaboration}{\textsc{Geant4}
  Collaboration}), \bibinfo{journal}{Nucl. Instrum. Methods Phys. Res., Sect.
  A} \textbf{\bibinfo{volume}{506}}, \bibinfo{pages}{250}
  (\bibinfo{year}{2003}).

\bibitem[{\citenamefont{Kuhr et~al.}(2019)\citenamefont{Kuhr, Pulvermacher,
  Ritter, Hauth, and Braun}}]{basf2}
\bibinfo{author}{\bibfnamefont{T.}~\bibnamefont{Kuhr}},
  \bibinfo{author}{\bibfnamefont{C.}~\bibnamefont{Pulvermacher}},
  \bibinfo{author}{\bibfnamefont{M.}~\bibnamefont{Ritter}},
  \bibinfo{author}{\bibfnamefont{T.}~\bibnamefont{Hauth}}, \bibnamefont{and}
  \bibinfo{author}{\bibfnamefont{N.}~\bibnamefont{Braun}}
  (\bibinfo{collaboration}{Belle II Framework Software Group}),
  \bibinfo{journal}{Comput. Software Big Sci.} \textbf{\bibinfo{volume}{3}},
  \bibinfo{pages}{1} (\bibinfo{year}{2019}).

\bibitem[{\citenamefont{{The Belle II Collaboration}}()}]{basf2-zenodo}
\bibinfo{author}{\bibnamefont{{The Belle II Collaboration}}},
  \emph{\bibinfo{title}{{\rm{Belle II Analysis Software Framework (basf2)}}}},
  \bibinfo{howpublished}{{https://doi.org/10.5281/zenodo.5574115}}.

\bibitem[{\citenamefont{Abudin{\'e}n et~al.}(2022)}]{punzi-loss}
\bibinfo{author}{\bibfnamefont{F.}~\bibnamefont{Abudin{\'e}n}}
  \bibnamefont{et~al.}, \bibinfo{journal}{Eur. Phys. J. C}
  \textbf{\bibinfo{volume}{82}}, \bibinfo{pages}{121} (\bibinfo{year}{2022}).

\bibitem[{\citenamefont{Punzi}(2003)}]{Punzi:2003bu}
\bibinfo{author}{\bibfnamefont{G.}~\bibnamefont{Punzi}},
  \bibinfo{journal}{eConf} \textbf{\bibinfo{volume}{C 030908}},
  \bibinfo{pages}{MODT002} (\bibinfo{year}{2003}).

\bibitem[{\citenamefont{Brandt et~al.}(1964)\citenamefont{Brandt, Peyrou,
  Sosnowski, and Wroblewski}}]{BRANDT196457}
\bibinfo{author}{\bibfnamefont{S.}~\bibnamefont{Brandt}},
  \bibinfo{author}{\bibfnamefont{C.}~\bibnamefont{Peyrou}},
  \bibinfo{author}{\bibfnamefont{R.}~\bibnamefont{Sosnowski}},
  \bibnamefont{and}
  \bibinfo{author}{\bibfnamefont{A.}~\bibnamefont{Wroblewski}},
  \bibinfo{journal}{Phys. Lett.} \textbf{\bibinfo{volume}{12}},
  \bibinfo{pages}{57} (\bibinfo{year}{1964}), ISSN \bibinfo{issn}{0031}.

\bibitem[{\citenamefont{Farhi}(1977)}]{PhysRevLett.39.1587}
\bibinfo{author}{\bibfnamefont{E.}~\bibnamefont{Farhi}},
  \bibinfo{journal}{Phys. Rev. Lett.} \textbf{\bibinfo{volume}{39}},
  \bibinfo{pages}{1587} (\bibinfo{year}{1977}).

\bibitem[{\citenamefont{Cowan et~al.}(2011)\citenamefont{Cowan, Cranmer, Gross,
  and Vitells}}]{cowan2011}
\bibinfo{author}{\bibfnamefont{G.}~\bibnamefont{Cowan}},
  \bibinfo{author}{\bibfnamefont{K.}~\bibnamefont{Cranmer}},
  \bibinfo{author}{\bibfnamefont{E.}~\bibnamefont{Gross}}, \bibnamefont{and}
  \bibinfo{author}{\bibfnamefont{O.}~\bibnamefont{Vitells}},
  \bibinfo{journal}{Eur. Phys. J. C} \textbf{\bibinfo{volume}{71}},
  \bibinfo{pages}{1554} (\bibinfo{year}{2011}).

\bibitem[{\citenamefont{Heinrich et~al.}(2021)\citenamefont{Heinrich, Feickert,
  Stark, and Cranmer}}]{Heinrich2021}
\bibinfo{author}{\bibfnamefont{L.}~\bibnamefont{Heinrich}},
  \bibinfo{author}{\bibfnamefont{M.}~\bibnamefont{Feickert}},
  \bibinfo{author}{\bibfnamefont{G.}~\bibnamefont{Stark}}, \bibnamefont{and}
  \bibinfo{author}{\bibfnamefont{K.}~\bibnamefont{Cranmer}},
  \bibinfo{journal}{J. Open Source Software} \textbf{\bibinfo{volume}{6}},
  \bibinfo{pages}{2823} (\bibinfo{year}{2021}).

\bibitem[{\citenamefont{Crivellin and Hoferichter}(2022)}]{Crivellin:2022gfu}
\bibinfo{author}{\bibfnamefont{A.}~\bibnamefont{Crivellin}} \bibnamefont{and}
  \bibinfo{author}{\bibfnamefont{M.}~\bibnamefont{Hoferichter}}
  (\bibinfo{year}{2022}), \eprint{arXiv:2211.12516}.

\bibitem[{\citenamefont{Gross and Vitells}(2010)}]{Gross:2010qma}
\bibinfo{author}{\bibfnamefont{E.}~\bibnamefont{Gross}} \bibnamefont{and}
  \bibinfo{author}{\bibfnamefont{O.}~\bibnamefont{Vitells}},
  \bibinfo{journal}{Eur. Phys. J. C} \textbf{\bibinfo{volume}{70}},
  \bibinfo{pages}{525} (\bibinfo{year}{2010}).

\bibitem[{\citenamefont{Bellini et~al.}(2011)}]{PhysRevLett.107.141302}
\bibinfo{author}{\bibfnamefont{G.}~\bibnamefont{Bellini}} \bibnamefont{et~al.}
  (\bibinfo{collaboration}{Borexino Collaboration}), \bibinfo{journal}{Phys.
  Rev. Lett.} \textbf{\bibinfo{volume}{107}}, \bibinfo{pages}{141302}
  (\bibinfo{year}{2011}).

\bibitem[{\citenamefont{Kamada et~al.}(2018)\citenamefont{Kamada, Kaneta,
  Yanagi, and Yu}}]{Kamada2018}
\bibinfo{author}{\bibfnamefont{A.}~\bibnamefont{Kamada}},
  \bibinfo{author}{\bibfnamefont{K.}~\bibnamefont{Kaneta}},
  \bibinfo{author}{\bibfnamefont{K.}~\bibnamefont{Yanagi}}, \bibnamefont{and}
  \bibinfo{author}{\bibfnamefont{H.-B.} \bibnamefont{Yu}}, \bibinfo{journal}{J.
  High Energy Phys.} \textbf{\bibinfo{volume}{06}}, \bibinfo{pages}{117}
  (\bibinfo{year}{2018}).

\bibitem[{sup()}]{supplemental}
\bibinfo{note}{See Supplemental Material at {[URL]} for additional plots.}

\end{thebibliography}

\clearpage




\section*{Supplementary information}
\renewcommand{\thefigure}{S\arabic{figure}}


\onecolumngrid

\setcounter{figure}{0}

This material is submitted as supplementary information for the Electronic Physics Auxiliary Publication Service

We provide a text file with numerical results for the observed 90\% CL upper limit on the cross section of $e^+e^- \to \mu^+\mu^- Z^{\prime}$, with $Z^{\prime} \to \text{invisible}$ as well as of the observed 90\% CL upper limit on $g^\prime$ as functions of $M_{Z^\prime}$.

\begin{figure}[!htb]
  \centering
  \includegraphics[width=0.70\columnwidth]{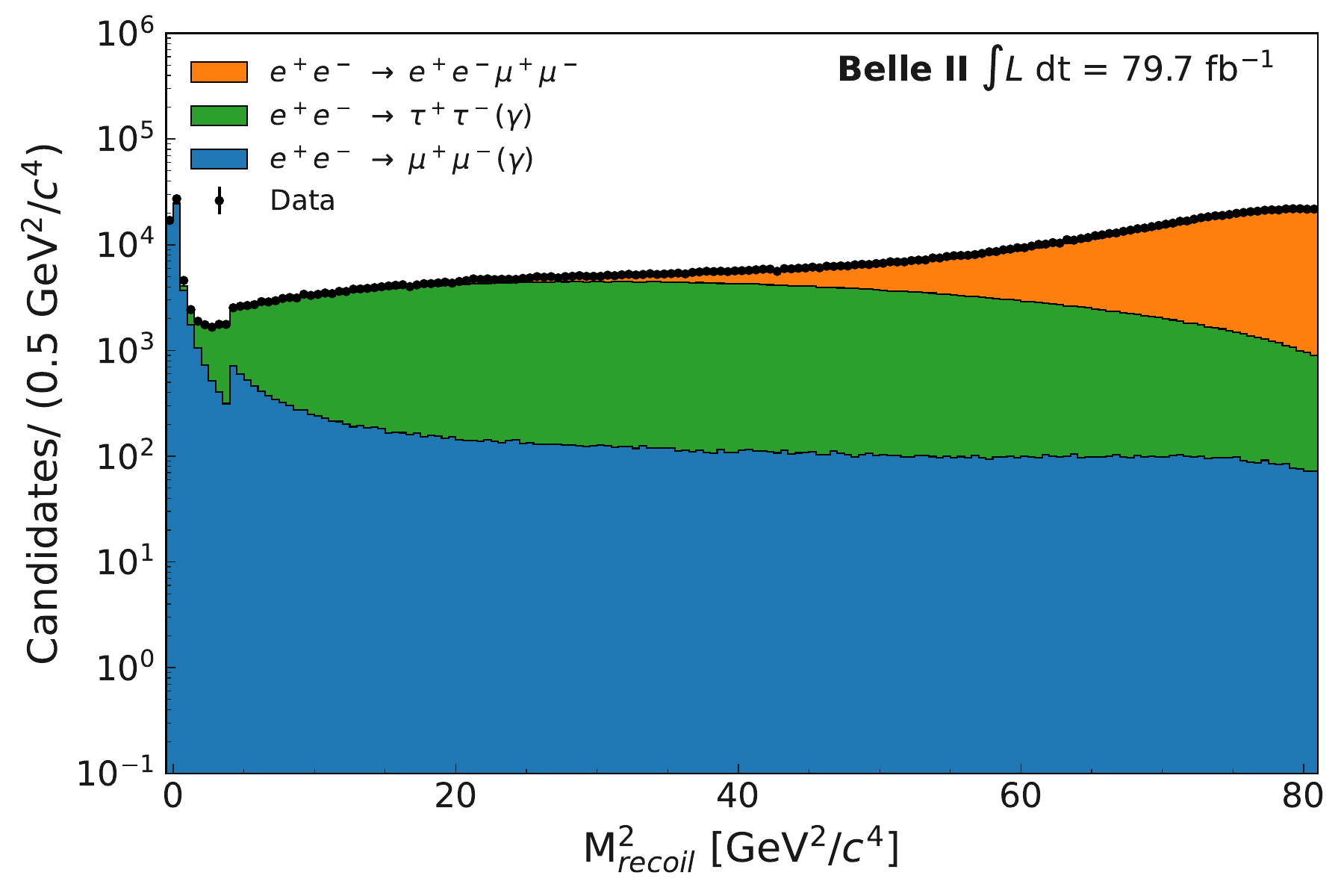}
  \caption{Squared recoil mass spectrum for the \mumu sample before the Punzi-net selection, compared with the stacked contributions from the various simulated background samples normalized to the integrated luminosity.
  }
  \label{fig:before_punzi}
\end{figure}

\begin{figure}[!htb]
  \centering
  \includegraphics[width=0.60\columnwidth]{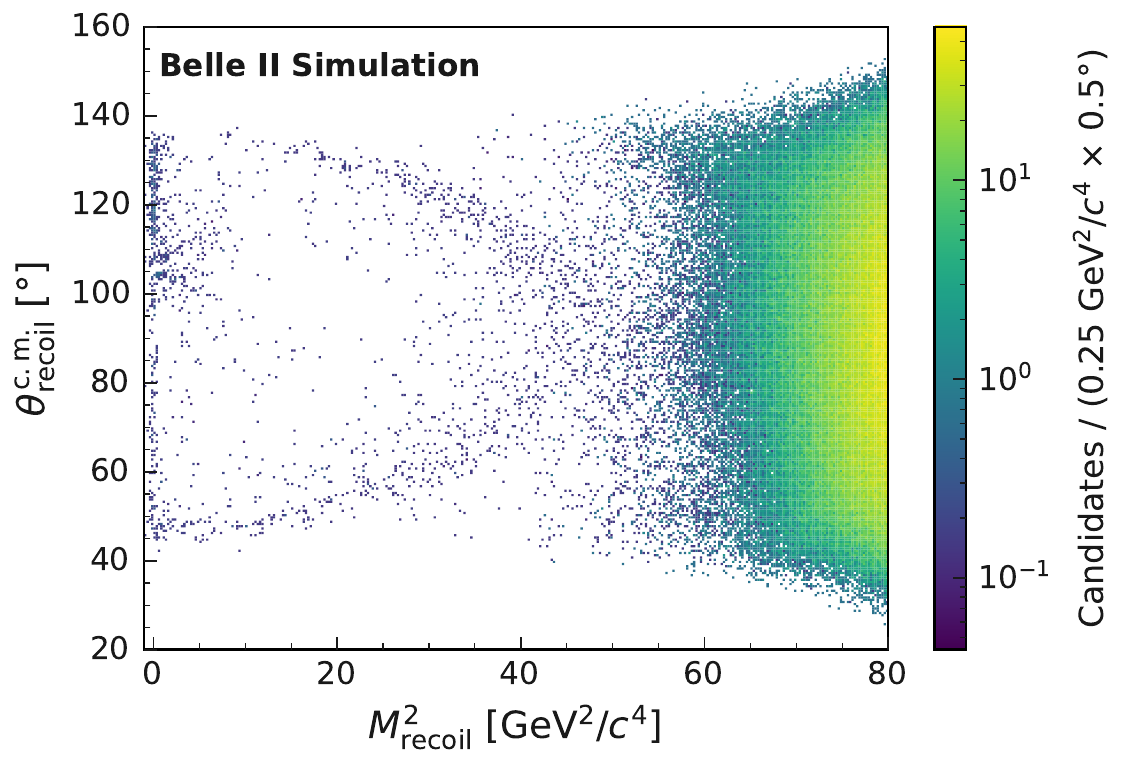}
  
  \includegraphics[width=0.60\columnwidth]{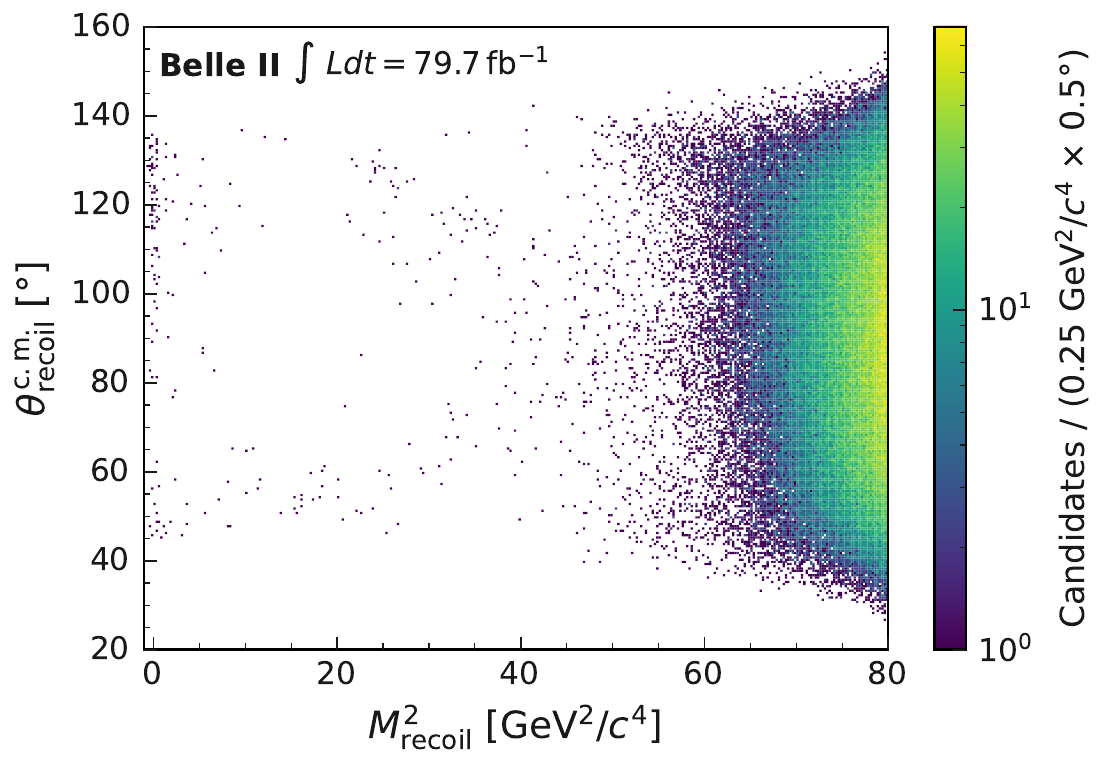}
  \caption{ Distribution of (top) expected background events and (bottom) data across the \treccm\  versus $M^2_{\rm{recoil}}$ plane after all the analysis selections.
  }
  \label{fig:after_punzi_2d}
\end{figure}

\begin{figure}[!htb]
  \centering
  \includegraphics[width=0.70\columnwidth]{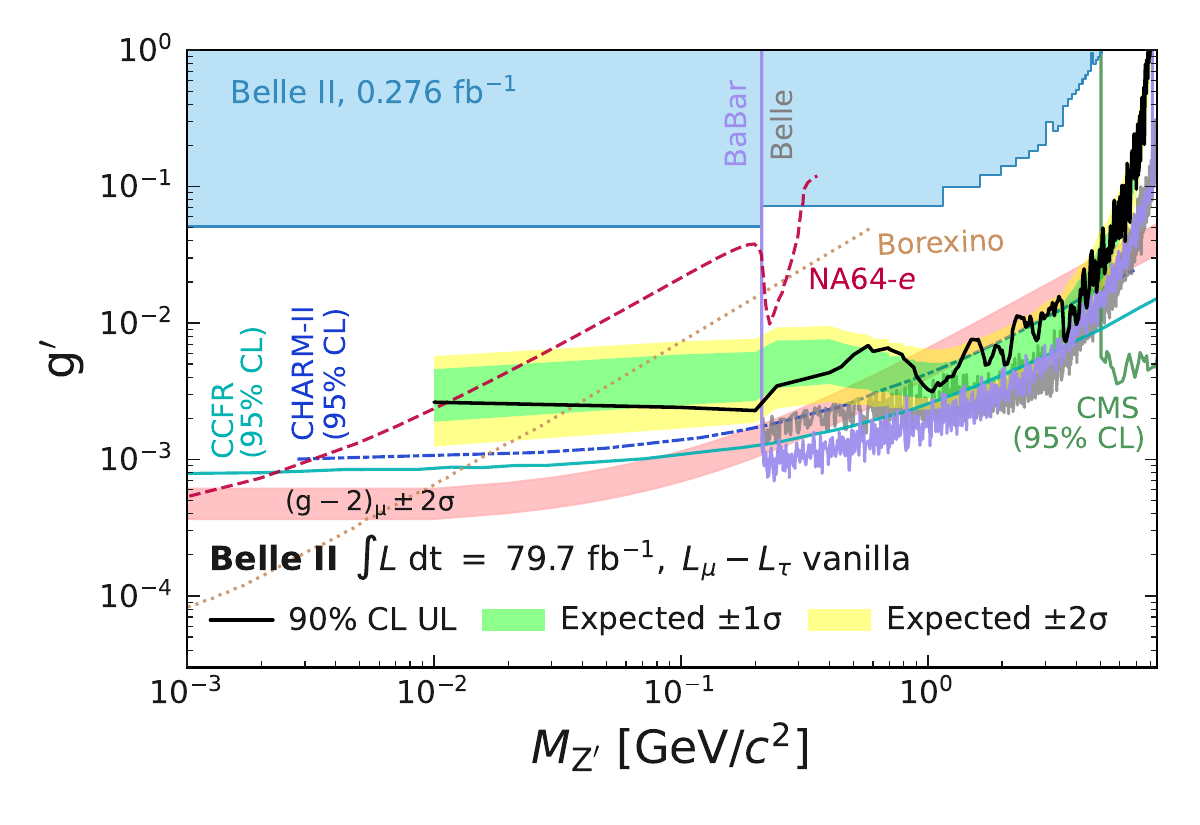}
  \caption{
  Observed 90$\%$ CL upper limits on the coupling   $g^{\prime}$ for the \lmultau\ vanilla model as functions of $M_{Z'}$ on a logarithmic scale. Existing limits from   BaBar~\cite{TheBABAR:2016rlg}, Belle~\cite{Belle:2021feg}, CMS~\cite{cms} (95\% CL), NA64-$e$~\cite{Andreev:2022txy}, and Belle~II~\cite{PhysRevLett.124.141801} are shown, along with constraints (95\% CL) derived from the trident production in neutrino experiments~\cite{PhysRevLett.113.091801, PhysRevLett.107.141302, Kamada2018}. The red band shows the region that could explain the muon anomalous magnetic moment $(g-2)_{\mu} \pm 2\sigma$~\cite{PhysRevLett.126.141801}.}
  \label{fig:gprime_lmultau_xlog}
\end{figure}

\begin{figure}[!htb]
  \centering
  \includegraphics[width=0.70\columnwidth]{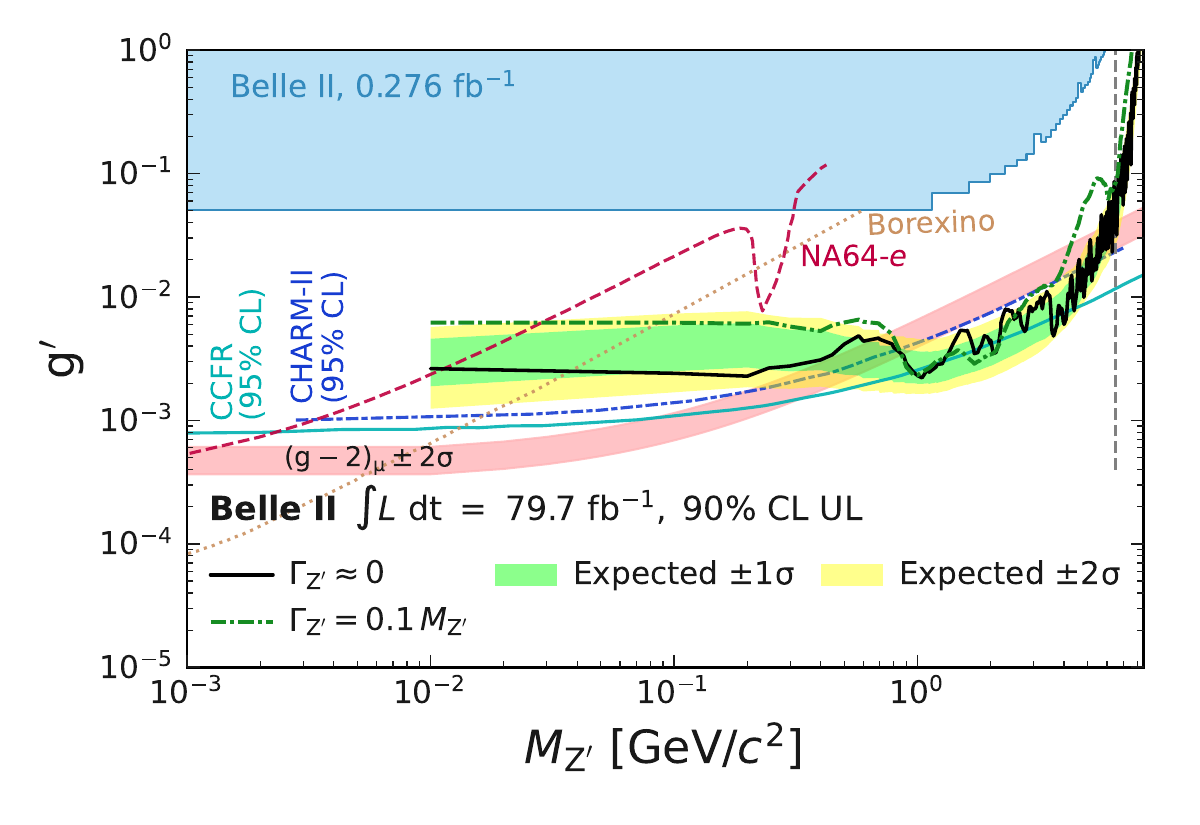}
  \caption{Observed 90$\%$ CL upper limits on the coupling  $g^{\prime}$ for the \lmultau\ fully invisible model as functions of $M_{Z'}$ in logarithmic scale.  Existing limits from NA64-$e$~\cite{Andreev:2022txy} and Belle~II~\cite{PhysRevLett.124.141801} are shown, along with constraints (95\% CL) derived from the trident production in neutrino experiments~\cite{PhysRevLett.113.091801, PhysRevLett.107.141302, Kamada2018}.
  The vertical dashed line indicates the limit beyond which the hypothesis $\cal{B}\rm(Z^{\prime} \to \chi \bar{\chi}) \approx 1$ is not valid in the negligible $\Gamma_{Z'}$ case.
  The red band shows the region that could explain the muon anomalous magnetic moment $(g-2)_{\mu} \pm 2\sigma$~\cite{PhysRevLett.126.141801}.}
  \label{fig:gprime_only_inv_xlog}
\end{figure}

We also show the limits on $g^\prime$ as functions of $M_{Z'}$ on a logarithmic scale in Fig.~\ref{fig:gprime_lmultau_xlog} for the \lmultau\ vanilla model and in  Fig.~\ref{fig:gprime_only_inv_xlog} for the \lmultau\ fully invisible model. 

\begin{figure}[!htb]
  \centering
  \includegraphics[width=0.70\columnwidth]{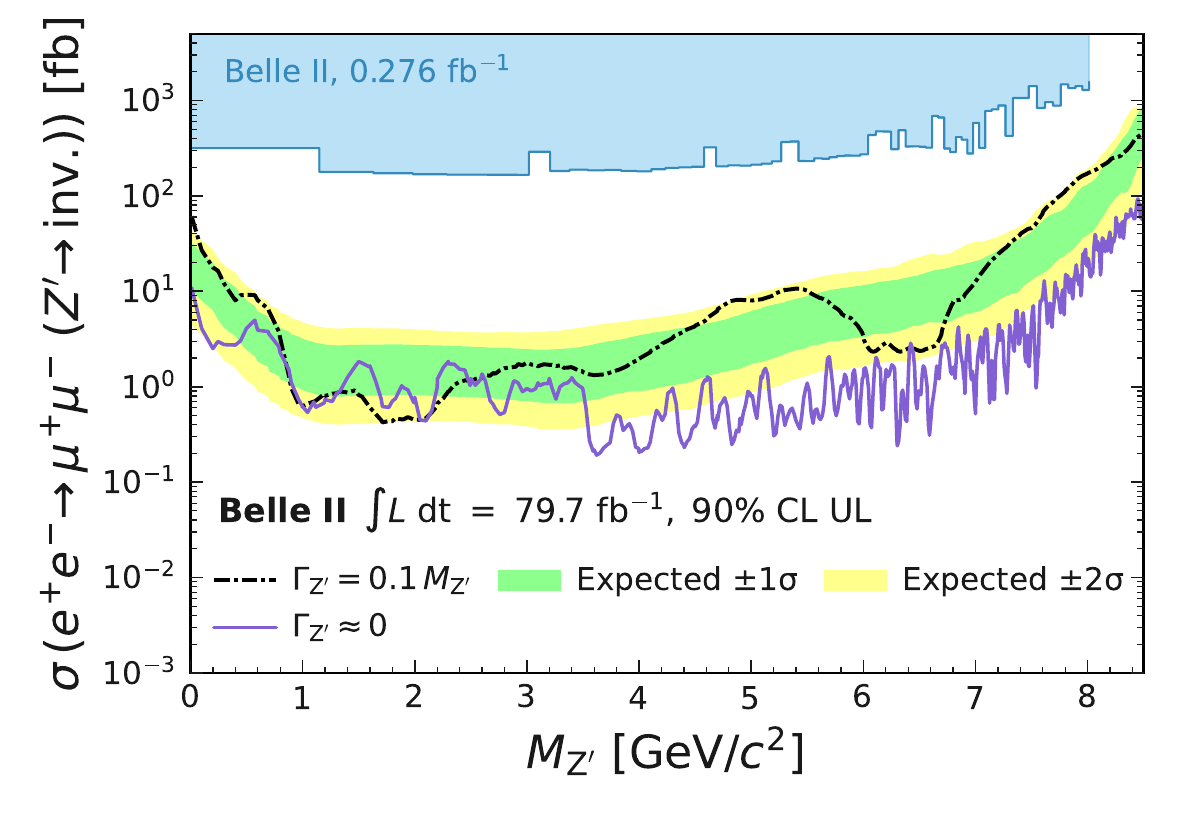}
  \caption{Observed 90$\%$ CL upper limits on the cross section $\sigma({e^+e^- \to \mu^+\mu^-Z^\prime,\,\,Z^\prime\to\rm{invisible}})$ as functions of the \zprime\ mass for $\Gamma_{Z'}=0.1 M_{Z'}$, including the $\pm 1$ and $\pm 2\sigma$ bands around the expected limits.
  Also shown are previous limits from Belle~II~\cite{PhysRevLett.124.141801} and the observed limits for the negligible width case.}
  \label{fig:cross_section_gamma_band}
\end{figure}

\begin{figure}[!htb]
  \centering
  \includegraphics[width=0.70\columnwidth]{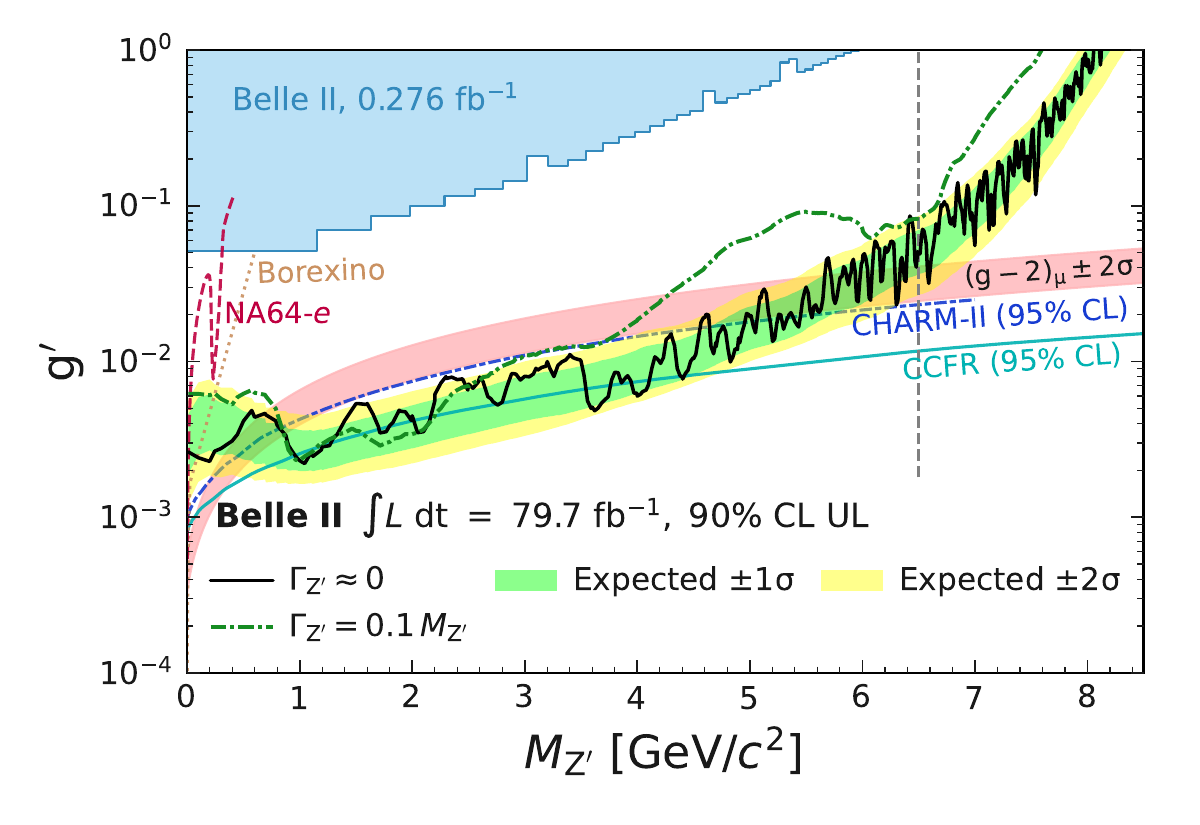}
  \caption{Observed 90$\%$ CL upper limits on the coupling   $g^{\prime}$ for the \lmultau\ fully invisible model as functions of $M_{Z'}$.  Existing limits from NA64-$e$~\cite{Andreev:2022txy} and Belle~II~\cite{PhysRevLett.124.141801} are shown, along with constraints (95\% CL) derived from the trident production in neutrino experiments~\cite{PhysRevLett.113.091801, PhysRevLett.107.141302, Kamada2018}.
  The vertical dashed line indicates the limit beyond which the hypothesis $\cal{B}\rm(Z^{\prime} \to \chi \bar{\chi}) \approx 1$ is not valid in the negligible $\Gamma_{Z'}$ case.
  The red band shows the region that could explain the muon anomalous magnetic moment $(g-2)_{\mu} \pm 2\sigma$~\cite{PhysRevLett.126.141801}.}
  \label{fig:gprime_intervals_invisible_with_neutrino}
\end{figure}

\begin{figure}[!htb]
  \centering
  \includegraphics[width=0.70\columnwidth]{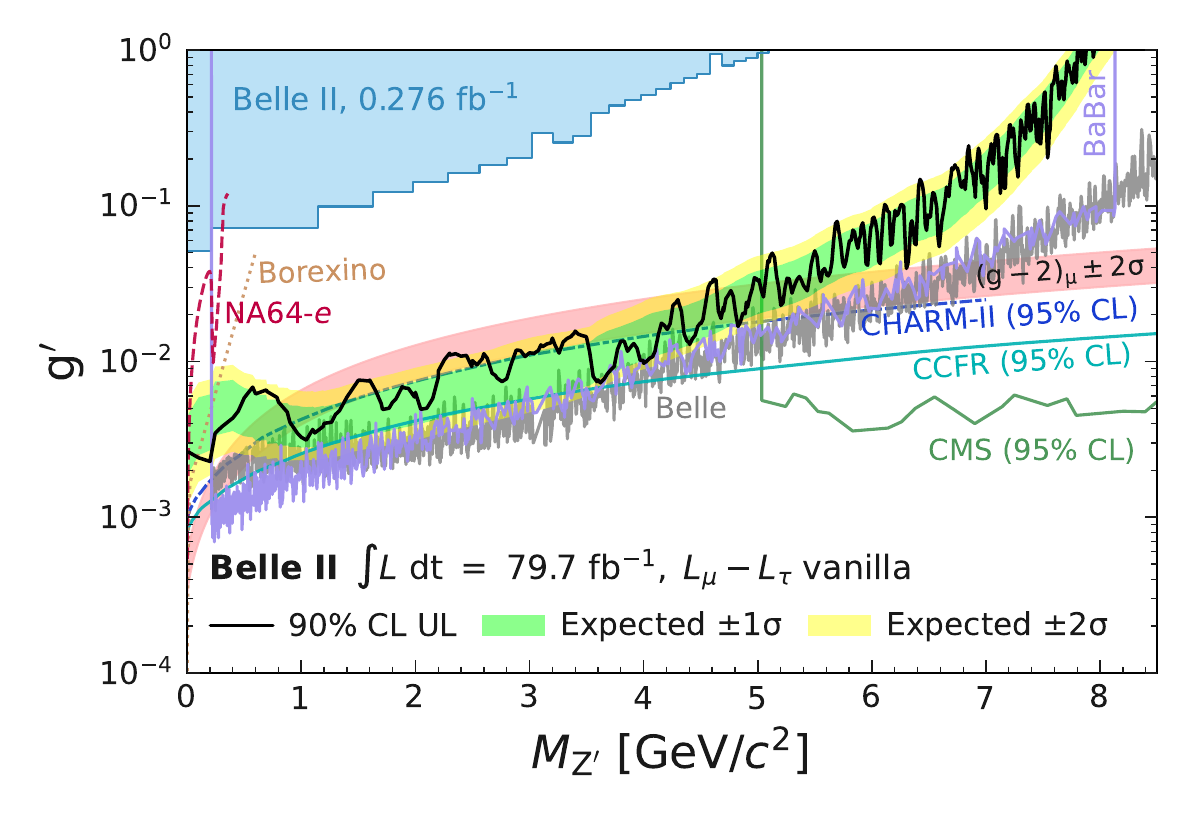}
  \caption{Observed 90$\%$ CL upper limits on the coupling   $g^{\prime}$ for the \lmultau\ vanilla model as functions of $M_{Z'}$. Existing limits from  BaBar~\cite{TheBABAR:2016rlg},  Belle~\cite{Belle:2021feg},  CMS~\cite{cms} (95\% CL), NA64-$e$~\cite{Andreev:2022txy}, and Belle~II~\cite{PhysRevLett.124.141801} are shown, along with constraints (95\% CL) derived from the trident production in neutrino experiments~\cite{PhysRevLett.113.091801, PhysRevLett.107.141302, Kamada2018}. The red band shows the region that could explain the muon anomalous magnetic moment $(g-2)_{\mu} \pm 2\sigma$~\cite{PhysRevLett.126.141801}.}
  \label{fig:gprime_intervals_lmu_ltau_with_neutrino}
\end{figure}

\clearpage


\end{document}